 \documentclass[review,12pt]{elsarticle}
 \usepackage{caption}
 \usepackage{placeins}
 \usepackage{graphicx}

\usepackage{subcaption}

\usepackage{latexsym}
\usepackage{amsmath}
\usepackage{amsfonts}
\usepackage{amssymb}
\usepackage{tikz}

\DeclareMathOperator{\arctantwo}{atan2}

\usetikzlibrary{trees}

\usepackage{comment}
\usepackage{enumitem}
\usepackage{booktabs}
\usepackage{wrapfig}
\usepackage{lscape}
\usepackage{multimedia}
\usepackage{stmaryrd} 

\usepackage{multirow}
\usepackage{url}
\usepackage{xcolor}
\definecolor{newcolor}{rgb}{.8,.349,.1}

\journal{Computers \& Fluids}

\begin{document}

\graphicspath{{pictures/}}

\begin{frontmatter}

\title{A mass-spring fluid-structure interaction solver: \\ application to flexible revolving wings}

\author[1]{Hung Truong}
\ead{dinh-hung.truong@univ-amu.fr}
\author[2]{Thomas Engels}
\ead{thomas.engels@ens.fr}
\author[3]{Dmitry Kolomenskiy}
\ead{dkolomenskiy@jamstec.go.jp}
\author[1]{Kai Schneider}
\ead{kai.schneider@univ-amu.fr}
\address[1]{Aix-Marseille Universit\'e, CNRS, Centrale Marseille, I2M UMR 7373, Marseille, 39 rue Joliot-Curie, 13451 Marseille Cedex 20, France}
\address[2]{LMD-IPSL, \'Ecole Normale Sup\'erieure-PSL, 24 Rue Lhomond, 75231 Paris Cedex 05, France}
\address[3]{Japan Agency for Marine-Earth Science and Technology (JAMSTEC), 3173-25 Showa-machi, Kanazawa-ku, Yokohama Kanagawa 236-0001, Japan}

\begin{abstract}
The secret to the spectacular flight capabilities of flapping insects lies in their wings, which are often approximated as flat, rigid plates. Real wings are however delicate structures, composed of veins and membranes, and can undergo significant deformation. In the present work, we present detailed numerical simulations of such deformable wings. Our results are obtained with a fluid--structure interaction solver, coupling a mass--spring model for the flexible wing with a pseudo-spectral code solving the incompressible Navier--Stokes equations. We impose the no-slip boundary condition through the volume penalization method; the time-dependent complex geometry is then completely described by a mask function. This allows solving the governing equations of the fluid on a regular Cartesian grid. Our implementation for massively parallel computers allows us to perform high resolution computations with up to 500 million grid points. The mass--spring model uses a functional approach, thus modeling the different mechanical behaviors of the veins and the membranes of the wing. We perform a series of numerical simulations of a flexible revolving bumblebee wing at a Reynolds number $Re=1800$. In order to assess the influence of wing flexibility on the aerodynamics, we vary the elasticity parameters and study rigid, flexible and highly flexible wing models. Code validation is carried out by computing classical benchmarks.
\end{abstract}


\end{frontmatter}

\section{Introduction}
A fundamental characteristics of insect flight are flexible wings, which play an important role for their aerodynamics \cite{AeroYoung,FSINakata2012,CombesI}, requiring lower forces than their rigid counter parts and producing reduced sound. Numerical simulation of insect flight is itself a sophisticated task, because it involves the solution of fluid-solid interaction problems. Thus, we have to model simultaneously the fluid part and the solid part by using two coupled solvers. 

The fluid solver \cite{Flusi} we are using in the present work has been developed previously and is called FLUSI \footnote{FLUSI: freely available for noncommercial use from GitHub (https://github.com/pseudospectators/FLUSI).}, a fully parallel software dedicated for modeling three-dimensional flapping flight in viscous flows. The heart of this software is the Fourier pseudospectral method with adaptive time stepping used for the discretization of the 3D incompressible Navier--Stokes equations. Moreover, the volume penalization method is used to take into account the no-slip boundary conditions on the interfaces between the fluid and the solid part. 

In \cite{KES14}, 
we performed numerical simulations of rotating triangular rigid wings at Reynolds number $Re=250$ to investigate the Leading-Edge Vortices (LEVs) as a function of the wing aspect ratio and the angle of attack.
High resolution direct numerical simulations of rotating and flapping bumblebee wings were presented in \cite{RigidRevolWing} using likewise the FLUSI code with rigid wings focusing again on the role of LEVs and the associated helicity production. 

The interaction of flapping bumblebees with turbulent inflow in free and tethered flight was studied in
\cite{EKSFLS19} using once again massively parallel computations with FLUSI. We found that the fluctuations of aerodynamic observables significantly grow with increasing turbulence intensity, even if the mean values are almost unaffected. Changing the length scale of the turbulent inflow, while keeping the turbulence intensity fixed, showed that the fluctuation level of forces and moments can be significantly reduced. We also found that the scale-dependent energy distribution in the surrounding turbulent flow is a relevant factor conditioning how flying insects control their body orientation. 

Nevertheless, a solver for solving the deformation of the structure was not fully developed in this software FLUSI. Hence, all previous simulations of insect flight have been performed with the essential assumption that the insect is composed of linked rigid parts including the wings.

In the current work we aim at investigating the role of wing elasticity on the flight performance. Consequently, a wing model is required for simulating the deformation of the solid part under the impact from the fluid. There are many models based on continuum mechanics theory, which are used in many well-known solid solvers. Among these, Finite Element Methods (FEM) are mostly used in both research and industrial fields due to their reliability and effectiveness. Despite this dominance, the requirement for faster and more efficient methods motivates the development of alternative models. One of these is the mass-spring system, which is known for its computational efficiency and straightforward implementation \cite{Jar12}. As part of this work, a solver using a network of masses and springs is developed to model a flexible insect wing and coupled with FLUSI.  

The story of mass-spring models started back at the end of the $20^{th}$ century when people were dealing with observable deformations of flexible objects in computer graphics applications, such as soft tissue, skin, hair, ball, cloth, textiles, etc. Being considered as one of the pioneers on this field, Terzopoulos et al. \cite{MembraneTerzopoulos} proposed using elasticity theory for modeling deformable objects instead of previously-used kinematic models, where the motion and the deformation of materials were prescribed. During this period, most of simulations of flexible objects had been calculated using finite element methods until the requirement of faster models was claimed by Eischen et al.~\cite{MembraneEischen}. Consequently, the development of physically based deformable models started to grow, especially in the field of computer graphics \cite{Cai16} and biomedical engineering \cite{Jar12}. While classic solvers, based on finite elements or finite difference methods, are generally employed in the static case for computing stress and strain in a structure, new solvers for deformable objects must have the ability to deal with large deformations and large displacements, i.e. the nonlinear regime. Furthermore, these models need to be fast enough since they are usually employed for real-time applications or coupled with other models which are already time-consuming. Among all the deformable models proposed, mass-spring networks stand out as the most intuitive and simplest due to their computational efficiency~\cite{DeforModelNealen}. Mass-spring systems have already been employed in many fields such as medical applications \cite{Jar12} (muscle, red blood cells and virtual surgery), computer graphics, fluid-structure interaction and insect flight. 
Miller and Peskin~\cite{PeskinClapandFling} used mass-spring networks to model insect wings in two-dimensional numerical simulations. A mass-spring network was used by Yeh and Alexeev~\cite{FSIYeh2016} to model a flexible plate swimmer and performed fluid-structure interaction simulation with Lattice-Boltzmann methods in 3D. The development of our solid solver is motivated by their mass-spring network approach, aiming to model the flexibility of insect wings in the three-dimensional case.  


The goal of this paper is to move from rigid to flexible wings and to present a fluid-structure interaction solver for flapping flight, based on the open access software FLUSI, where we integrated a solid model based on mass-spring systems. However, the wing kinematics of insects is very difficult to obtain, since the measurements usually require high tech equipment to capture all the dynamic motion at small time scale and length scale. Instead, a revolving wing model is usually employed to study the aerodynamics of flapping wings thanks to its simplicity. Accordingly, the flow fields and force generation aspects of revolving wings have been analyzed for a wide range of parameters, as reported in the literature \cite{RevolWingLentink09,RevolvWingJardin14,RevolWingJones16}. Di et al.~\cite{RevolWingDi18} studied the role of forewings in generating LEVs of three revolving insect wing models: hawkmoth (Manduca sexta), bumblebee (Bombus ignitus) and fruitfly (Drosophila melanogaster). Van de Meerendonk et al.~\cite{FlexInfluenceMeerendonk} investigated experimentally the flow field and fluid-dynamic loads of a flexible revolving wing to quantify the influence of flexibility on the force generation performance of the wing. In our study, we also consider revolving flexible bumblebee wings and assess the influence of the wing deformation.

The outline of this paper is the following: In sec.~\ref{sec:flex} we present a mass-spring model for describing the flexible insect wings.
The wing structure of the considered bumblebee wings and its mass distribution are detailed in sec.~\ref{sec:Wing_structure}.
The numerical artillery of fluid-structure interaction is explained in sec.~\ref{sec:FSI} and some validation tests are given.
The numerical results as well as the discussion about the influence of the flexibility on the aerodynamic performance are presented in sec.~\ref{sec:results}.
Finally, conclusions are drawn in sec.~\ref{sec:conclusions}, including some perspectives for future investigations.


\section{Flexible wing}
\label{sec:flex}

\subsection{Mass-spring model}
 
 The very basic idea of the mass-spring model is to discretize an object using mass points $m_i \, (i=1,...,n)$ connected by springs. There exist many kinds of springs for different purposes but in the limit of our work, we have used only linear extension and bending springs to model insect wings. The dynamic behavior of the mass-spring system, at a given time $t$, is defined by the position $\mathbf{x}_i$ and the velocity $\mathbf{v}_i$ of the mass point $i$. For this, we need to solve the dynamic equations of the system, which govern their motion in time under certain external forces. This is one of the elegant advantages of the mass-spring network where these governing equations are simply the corresponding classical well-known Newton's second law, given in eqns. (\ref{eqn:Newton_law}).

\begin{equation}
\begin{split}
\mathbf{F}_i & = m_i \mathbf{a}_i \\
\mathbf{F}_i & = \mathbf{F}^{int}_i + \mathbf{F}^{ext}_i \ \ \ \ \ \ \rm{for} \ \ i=1...n  \\
\mathbf{v}_i (t=0) & = \mathbf{v}_{0,i}\\
\mathbf{x}_i (t=0) & = \mathbf{x}_{0,i}
\end{split}
\label{eqn:Newton_law}
\end{equation}
where $n$ is the number of mass points, $\mathbf{F}_i$ is the total force (internal force $\mathbf{F}^{int}_i$ and external force $\mathbf{F}^{ext}_i$) acting on the $i^{th}$ mass point, $m_i$, $\mathbf{a}_i$ are mass and acceleration of the $i^{th}$ mass point, respectively.

Here, all terms are quite simple to derive except for the forces. The external forces come from outside of the system and depend on the surrounding field and the problem we are dealing with. On the other hand, the internal forces represent the restoring forces caused by the springs. The complicated properties of these forces make the system (\ref{eqn:Newton_law}) become nonlinear. Hence, we have a nonlinear system of $3n$ second order ordinary differential equations (ODEs) corresponding to three dimensions $x,y,z$ and $n$ mass points. In the general case, this system (\ref{eqn:Newton_law}) needs to be solved numerically since its analytical solution cannot be derived. In practice, it is more convenient to convert a system of $3n$ second order equations into a system of $6n$ first order equations by using the relations $\mathbf{a}_i = d\mathbf{v}_i/dt$ and $\mathbf{v}_i = d\mathbf{x}_i/dt$. Eqns. (\ref{eqn:Newton_law}) can then be rewritten as below:
\begin{equation}
\begin{split}
\frac{d \mathbf{x}_i}{dt} & = \mathbf{v}_i \\
m_i \frac{d \mathbf{v}_i}{dt} & = \mathbf{F}^{int}_i + \mathbf{F}^{ext}_i   \ \ \ \ \ \ \ \ \rm{for} \ \ i=1...n  \\
\mathbf{v}_i (t=0) & = \mathbf{v}_{0,i} \\
\mathbf{x}_i (t=0) & = \mathbf{x}_{0,i}
\end{split}
\label{eqn:Newton_law_first_order}
\end{equation}
Let us call $\mathbf{q} = \big[ \mathbf{x}_i, \ \ \mathbf{v}_i \big]^\intercal$ the phase vector containing positions and velocities of all mass points and $\mathbf{f} (\mathbf{q}) = \big[ \mathbf{v}_i, \ \ m_i^{-1} (\mathbf{F}^{int}_i + \mathbf{F}^{ext}_i) \big]^\intercal$ the right hand side function, eqns.(\ref{eqn:Newton_law_first_order}) can be rewritten again under the familiar form of a system of first order ODEs as follows:
\begin{equation}
\begin{split}
\frac{d \mathbf{q}}{dt} & =  \mathbf{f} (\mathbf{q}, \mathbf{t}) \\
\mathbf{q} (t=0) & = \mathbf{q_0}
\end{split}
\label{eqn:first_order_ODEs}
\end{equation}
For time stepping, eqns. (\ref{eqn:first_order_ODEs}) need to be discretized using an appropriate numerical scheme. The choice for this scheme is not trivial since it depends on many factors. Due to the size of the wings, the mass-spring network contains a lot of very small springs, which make the system very stiff and we need an implicit scheme for time marching. For this reason, either a centered scheme or a backward scheme can be used. Although centered schemes are usually in favor for their conservation property without numerical diffusion, a centered scheme, for example the trapezoidal scheme, can lead to numerical instability at some points because the eigenvalues of the operator of the time discretization lie exactly on the imaginary axis, the boundary of the stability zone \cite{ThomasThesis}. Furthermore, the coupling between the fluid and the solid solver will require an adaptive time stepping scheme. For all these reasons, a second order backward differentiation scheme with variable time steps \cite{BDFscheme} is used to discretize eqns. (\ref{eqn:first_order_ODEs}) in time.

\begin{equation}
\begin{split}
\mathbf{q}_i^{n+1} - \frac{(1+\xi)^2}{1+2\xi} \mathbf{q}_i^{n} + \frac{\xi^2}{1+2\xi} \mathbf{q}_i^{n-1} & = \frac{1+\xi}{1+2\xi} \Delta t^n \mathbf{f}(\mathbf{q}_i^{n+1}) \\
\end{split}
\label{eqn:discretized_ODEs}
\end{equation}
where $\xi=\Delta t^n / \Delta t^{n-1}$ is the ratio between the current $\Delta t^n$ and the previous time step $\Delta t^{n-1}$. Eqn. (\ref{eqn:discretized_ODEs}) is a system of nonlinear equations with the variable $\mathbf{q}^{n+1}$, the phase vector of the system at the current time step, which needs to be solved. The Newton--Raphson method, a powerful iterative method, is employed to solve this nonlinear system of equations. With an initial guess, which is reasonably close to the true root of the equations, Newton--Raphson helps to find approximations of the root with the rate of convergence estimated to be quadratic. For our mass-spring solver, the initial guess chosen is the phase vector $\mathbf{q}^{n}$ solved at the previous time step; this allows the Newton--Raphson method to converge quickly since the structure is advanced slowly and smoothly in time, which assures that $\mathbf{q}^{n+1}$ remains close to $\mathbf{q}^{n}$. In most cases, the Newton--Raphson method in the solver needs three to four iterations to converge within a relative or absolute $L_2$ norm error of $10^{-6}$ as the stopping criterion.

\subsection{Extension springs and bending springs}

\begin{figure}[ht]
\centering
  \begin{tabular}{@{}c@{}}
    \includegraphics[scale=0.35]{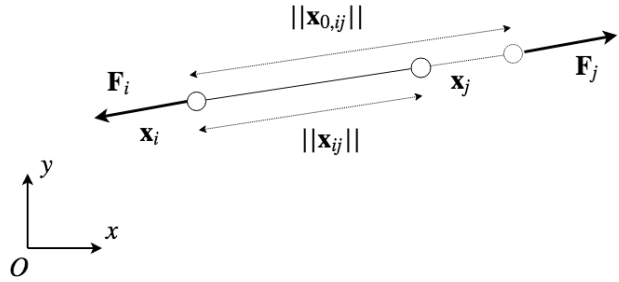} \\[\abovecaptionskip]
    \small (a) extension spring
  \end{tabular}

  \vspace{\floatsep}

  \begin{tabular}{@{}c@{}}
    \includegraphics[scale=0.35]{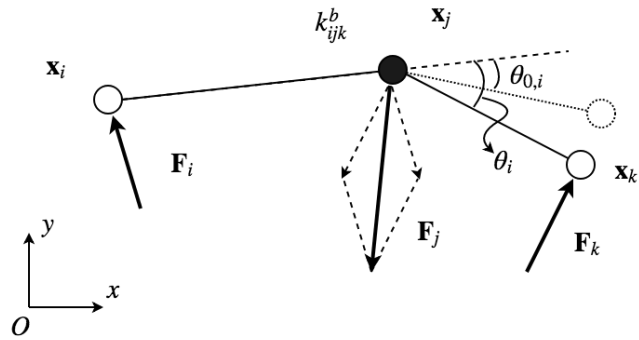} \\[\abovecaptionskip]
    \small (b) bending spring
  \end{tabular}

  \caption{Illustration of the restoring forces corresponding to the deformation of extension and bending springs. }\label{fig:springs}
\end{figure}

Extension springs (figure~\ref{fig:springs}a) and bending springs (figure~\ref{fig:springs}b) are common mechanical devices, which resist against the external forces to get back to their resting positions. The former is designed to operate with axial forces, while the latter is used for torques. The relations between the displacement and the restoring force are given by:

\begin{itemize}
\item Linear extension spring

\begin{equation}
\begin{split}
\mathbf{F}_{i} & = k^e_{ij} \left( ||\mathbf{x}_{ij}|| - ||\mathbf{x}_{0,ij}|| \right) \mathbf{e}_{ij} \\
\mathbf{F}_{j} & = - \mathbf{F}_{i}
\end{split}
\label{eqn:extension_spring}
\end{equation}
where $k^e_{ij}$ is the extension stiffness, $\mathbf{e}_{ij} = (\mathbf{x}_{j} - \mathbf{x}_{i}) / {|| \mathbf{x}_{j} - \mathbf{x}_{i} ||}$ is the unit position vector and $\mathbf{F}_{i}$ and $\mathbf{F}_{j}$ are the restoring forces of the extension spring acting on two points $i$ and $j$, respectively;

\item Linear bending spring 

\begin{equation}
\mathbf{M}_{ijk} = - k^b_{ijk} (\theta_{ijk} - \theta_{0,ijk}) \\
\label{eqn:bending_spring_moment}
\end{equation}

or in terms of forces

\begin{equation}
\begin{split}
\mathbf{F}_{i} & = k^b_{ijk} (\theta_{ijk} - \theta_{0,ijk}) \left( {\mathbf{e}_{ij}} \times {\mathbf{e}_{jk}} \right) \times {\mathbf{e}_{ij}} \\
\mathbf{F}_{k} & = k^b_{ijk} (\theta_{ijk} - \theta_{0,ijk}) \left( {\mathbf{e}_{ij}} \times {\mathbf{e}_{jk}} \right) \times {\mathbf{e}_{jk}} \\
\mathbf{F}_j & = - \mathbf{F}_{i} - \mathbf{F}_{k}
\label{eqn:bending_spring}
\end{split}
\end{equation}
where $\mathbf{M}_{ijk}$ is the restoring moment, $k^b_{ijk}$ is the bending stiffness, $\theta_{0,ijk}$ is the initial bending angle among three points $i,j$ and $k$, $\theta_{ijk}$ the current bending angle and $\mathbf{F}_{i}$, $\mathbf{F}_{j}$, $\mathbf{F}_{k}$ are the restoring force vectors (as shown in figure \ref{fig:springs}) of the bending spring acting on three points $i,j$ and $k$, respectively.

\end{itemize}

However, the calculation of $\theta_{ijk}$ in eqn.(\ref{eqn:bending_spring}) is not trivial since it involves the geometrical definition of the angle in 3D space with respect to a fixed coordinate system. Firstly, we consider a simpler case when the three points are in the same plane, a 2D coordinate system $Oxy$. This leads to $\mathbf{x}_{i}=(x_i, y_i)^T$, $\mathbf{x}_{j}=(x_j, y_j)^T$ and $\mathbf{x}_{k}=(x_k, y_k)^T$. The angle is now determined by:

\begin{equation}
\begin{split}
\theta_{ijk} = \arctantwo \Bigl[ & (y_k-y_j)(x_j-x_i)-(y_j-y_i)(x_k-x_j), \\ 
 & (x_k-x_j)(x_j-x_i)+(y_k-y_j)(y_j-y_i) \Bigr]
\label{eqn:theta_2D}
\end{split}
\end{equation}

Here, $\arctantwo$ (usually known as two-argument arctangent) is a special function first introduced in computer programming languages to give a correct and unambiguous value for the angle by taking into account the sign of both arguments. This function helps us to calculate on the whole space when the angle can vary in the range of $(-\pi,\pi]$, instead of the range of $(-\pi/2,\pi/2)$ when using the standard arctangent function.

\begin{figure}[th]
\centering
\includegraphics[width=0.45\textwidth]{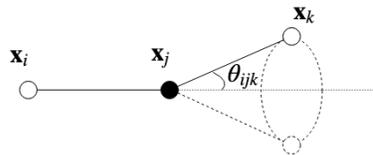}
\caption{All possible shapes of a bending spring corresponding to one bending angle $\theta_{ijk}$.}
\label{fig:one_angle_problem}
\end{figure}

For a problem in 3D space, only one bending angle will not be sufficient. This can be easily seen by considering a simple case of one bending spring. At an instant $t$, the spring is deformed and has a bending angle $\theta_{ijk}$. Nevertheless, corresponding to this $\theta_{ijk}$, there is an infinite number of solutions $\mathbf{x}_{i}$, $\mathbf{x}_{j}$ and $\mathbf{x}_{k}$ that can satisfy this deformation and the set of all these solutions forms a cone, like shown in figure \ref{fig:one_angle_problem}. Consequently, one more angle is needed to obtain a unique solution. To define these two bending angles, the same bending spring as the 2D case above is considered but now in a 3D coordinate system $Oxyz$. The bending spring is projected on the $Oxy$ and the $Oxz$ planes which gives us two 2D bending angles $\theta^y_{ijk}$ and $\theta^z_{ijk}$ on the $Oxy$ and the $Oxz$ planes, respectively. Then, like in the 2D problem, these two bending angles are calculated as below:

\begin{equation}
\begin{split}
\theta^y_{ijk} = \arctantwo \Bigl[&(x_j-x_i)(y_k-y_j) - (x_k-x_j)(y_j-y_i), \\
&(x_j-x_i)(x_k-x_j) + (y_j-y_i)(y_k-y_j) \Bigr] \\
\theta^z_{ijk} = \arctantwo \Bigl[&(x_j-x_i)(z_k-z_j) - (x_k-x_j)(z_j-z_i), \\
&(x_j-x_i)(x_k-x_j) + (z_j-z_i)(z_k-z_j) \Bigr]
\end{split}
\label{eqn:two_bending_angles}
\end{equation}

\subsection{Functional approach - vein and membrane models}
\label{Sec:functional_approach}

Modeling insect wings is challenging due to the fact that these wings have complex structures composed of a network of veins, partly connected through hinges, with thin membranes spanned in between and their elasticity properties are still poorly understood. Certain studies have shown that the vein arrangements in insect wings have strong impact on their mechanical properties \cite{CombesI,CombesII}. Thus, it will be inaccurate to consider a wing as a homogeneous structure; the vein pattern as well as the difference in terms of mechanical behaviors between vein and membrane need to be taken into account. However this is not an easy task, since insects are the most diverse group of animals living on Earth with more than one million known species with varying wing sizes and wing shapes. As a result, in this study, we want to limit ourself to a specific case when we examine only bumblebee (Bombus ignitus). Bumblebee wings are mainly composed of veins and membranes. The longitudinal veins extending along the wing in the spanwise direction are big, hollow and providing conduits for nerves while the cross veins are smaller, solid and connecting the longitudinal veins to form a kind of truss structure. In the space between these veins, we find a thin layer called membrane.

From a mechanical point of view, veins can be considered as rods which resist mainly the torsion and bending deformation. On the other hand, a membrane is fabric-like, it behaves like a piece of cloth which resists again the extension deformation. Consequently, instead of considering the wing as a homogeneous structure, a functional approach is used to distinguish veins and membranes. We then propose two models using mass-spring networks to imitate the mechanical behavior of the vein and the membrane.

\begin{figure}[th]
\centering
\includegraphics[width=0.45\textwidth]{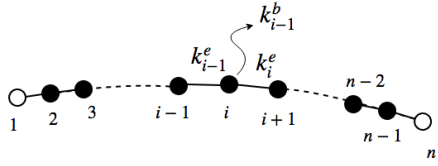}
\caption{Illustration of a vein modeled by mass points, extension springs and bending springs. White circles represent mass points, solid lines represent extension springs, black circles represent both mass points and bending springs.}
\label{fig:vein_model}
\end{figure}

A vein is considered as a rod whose length is much greater than its height and width. The total effect of all the external loads applied on a mechanical structure results in deformation which can usually be classified into three main types: bending, twisting and stretching. Although the whole wing is observed to be twisted significantly in many studies using high speed cameras or the digital particle image velocimetry \cite{WingDeformWalker1,WingDeformWalker2,WingDeformWalker3,WingDeformBomphrey,WingDeformMountcastle}, it is not entirely clear that torsion happens locally at veins or the unsynchronized bending deformations between veins cause the whole wing to twist. To simplify the model, we study only the latter in which we ignore the local torsion of veins and model solely the bending deformation of veins by using extension and bending springs. Thus, we model a vein by a curve line which is discretized by $n$ mass points as shown in figure \ref{fig:vein_model}. Two neighboring points are connected with each other by an extension spring (e.g. the mass points $i$ and $i+1$ are connected by the extension spring $k^e_i$) and three neighboring points are connected with each other by a bending spring (e.g. the mass points $i-1$, $i$ and $i+1$ are connected by the bending spring $k^b_{i-1}$).

When flapping, most of external forces will act in the direction perpendicular to the wing surface. As a result, the stretching deformation is negligible comparing to the bending deformation. Thus, the role of the extension springs in the model is solely to conserve the length of the vein. The stiffness of these extension springs is artificial and they do not need to reflect the mechanical property of the vein itself. They should be chosen stiff enough to make the rod unstretched but not too stiff to avoid problems with numerical stability when integrating the dynamical system in time.

\begin{figure}[th]
\centering
\includegraphics[width=0.45\textwidth]{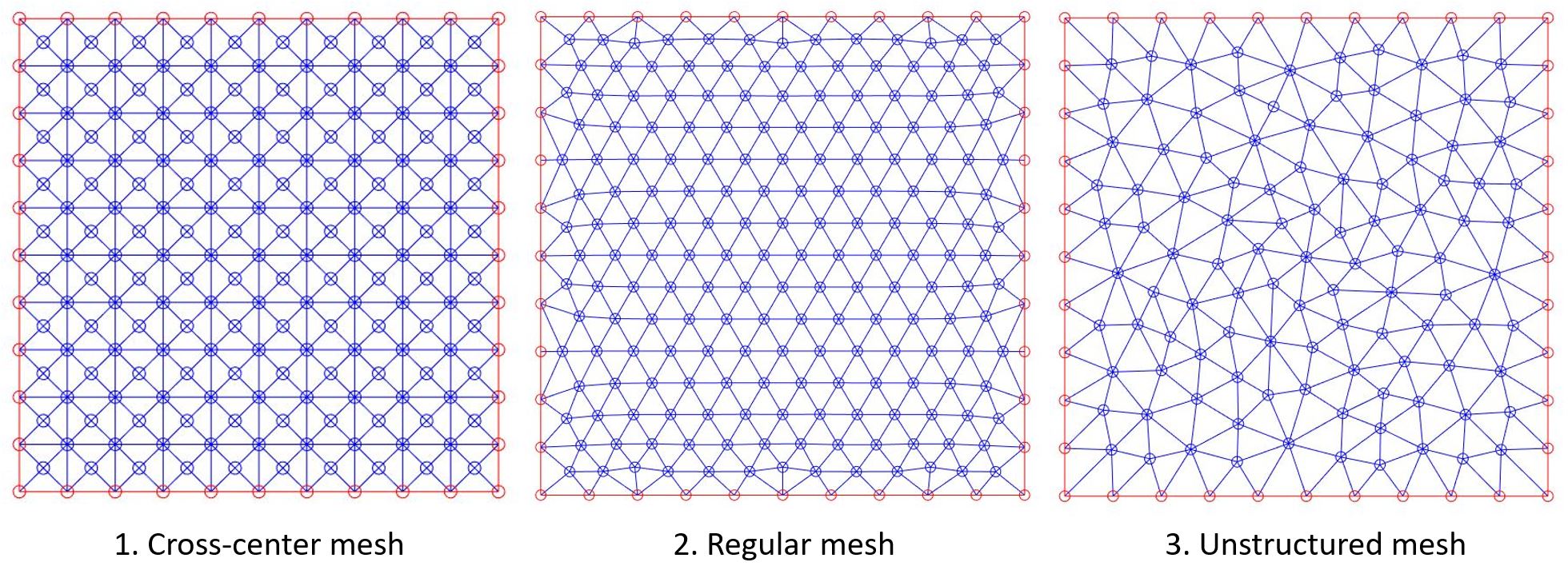}
\caption{Different mesh structures for 2D discretization \cite{MembraneOmori}.}
\label{fig:mesh_type_for_membrane}
\end{figure}

Compared to veins, membranes are totally different in terms of geometrical and mechanical properties. Geometrically, a membrane is an object whose thickness is much smaller than its extent. Consequently, a membrane is usually considered as a planar two-dimensional sheet or a set of planar sheets in the case of non-planar three-dimensional membranes \cite{NonplanarMembraneFenner,NonplanarMembraneWhite}. On the mechanical side, a membrane is a special kind of structure compared to other structural elements, i.e. a rod, a bar, a plate or a beam. It behaves like a piece of cloth which is much easier to be bent than to be stretched or compressed. Keeping these in mind, the membrane part of the wing is modeled by a 2D sheet which is discretized by a system of mass points and extension springs. There are several ways to discretize a 2D sheet (as shown in figure \ref{fig:mesh_type_for_membrane}) but an unstructured triangular mesh needs to be employed for our problem due to the complicated geometry of insect wing. Moreover, an unstructured mesh is preferred for modeling isotropic membranes \cite{MembraneBoyle} since the random orientations of the springs will average out the forces.

\subsubsection{Correlation between mass-spring network models and continuum constitutive laws}

Besides the mesh topology, the parameter setting is another challenge that one has to solve in order to correctly model the material of which the object is made. There are two main parameters needed to be assigned for a mass-spring model: the masses and the spring stiffness. Although a Voronoi diagram can be used to find the masses in a proper way \cite{MembraneDeussen}, selecting spring stiffness is still an open question and there are two common solutions to overcome this \cite{MSMLloyd}. The first approach is based on optimization methods to minimize the difference between the results solved by the mass-spring model and the reference data. These reference data can come from the measurements, the visual appearance of real objects \cite{MembraneLouchet} or numerical solutions using validated methods such as finite element methods \cite{MSMBianchiMesh,MSMBianchiParam}. In general, this approach cannot be applied if the system has too many degrees of freedom with many unknown spring constants or the mesh topology varies in time since one set of parameters works for solely one mesh structure. Otherwise, tuning the spring stiffness by using optimization can give satisfying results with reasonable computational cost.

\begin{figure}[th]
\centering
\includegraphics[width=0.45\textwidth]{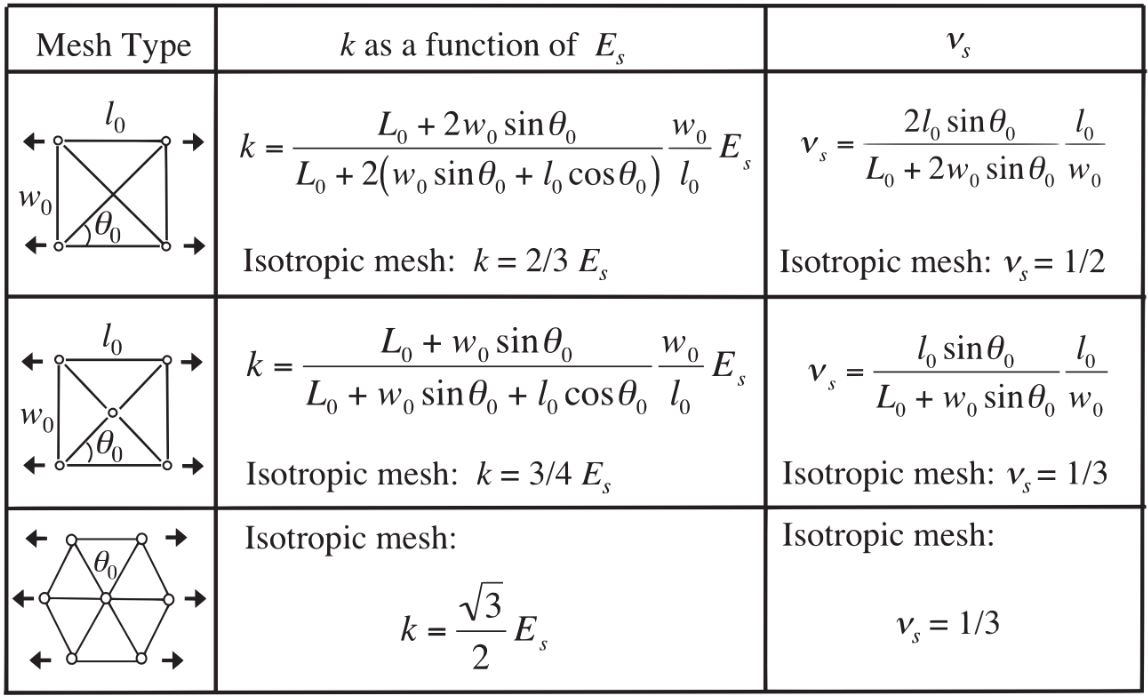}
\caption{Relationship between the spring constant $k$, Young modulus $E_s$, and Poisson ratio $\nu_s$ in the small deformation limit for a 2D membrane under uniaxial deformation. The membrane is discretized by three types of meshes and subjected to homogeneous deformation. Figure adapted from \cite{MembraneOmori}.}
\label{fig:Relation_k_Es}
\end{figure}

The second way is about deriving a relation between spring stiffness and other continuum mechanic properties, such as Young modulus, the Poisson ratio and the flexural rigidity. In contrast to the discrete models, the elasticity parameters have been obtained for many materials and can be used to calculate the corresponding spring stiffness. Omori et al. \cite{MembraneOmori} succeeded in doing this for a planar membrane by considering a 2D sheet under small uniaxial deformation. The relation between spring networks and continuum models for three types of meshes is shown in figure \ref{fig:Relation_k_Es}.

\begin{figure}[ht]
\centering
  \begin{tabular}{@{}c@{}}
     \includegraphics[width=.9\linewidth]{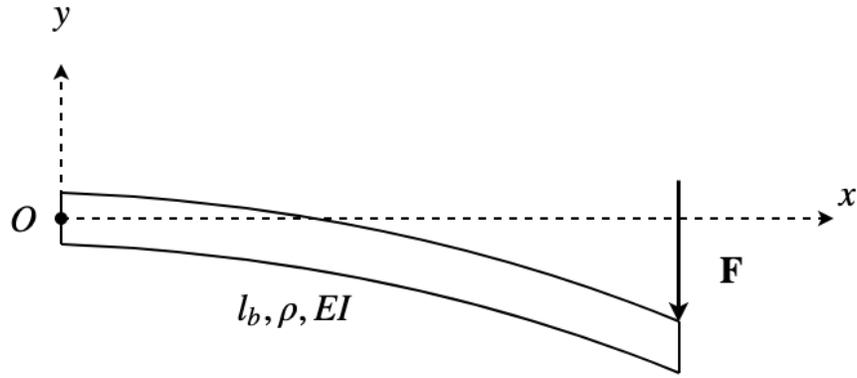}\\
    \small (a) Continuous beam
  \end{tabular}

  \begin{tabular}{@{}c@{}}
    \includegraphics[width=.9\linewidth]{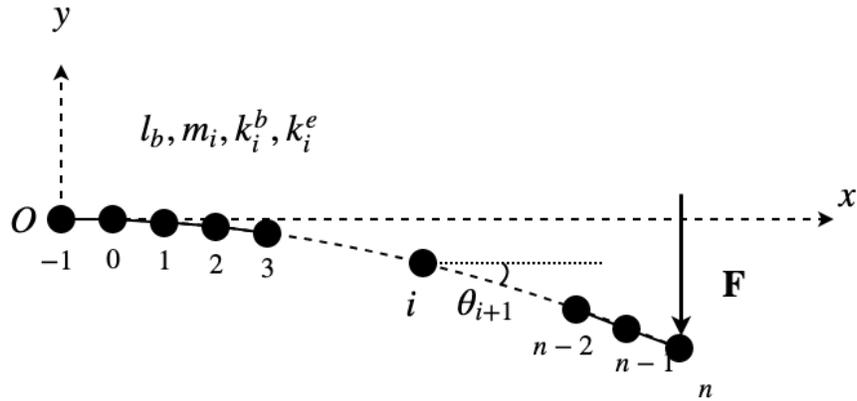}\\
    \small (b) Discrete mass-spring network
  \end{tabular}
  \caption{Illustration of deformation corresponding to forces applied on extension and bending springs.}
  \label{fig:cantilever_beam_point-force}
\end{figure}

For the vein model, a relation between the flexural rigidity $EI$ and the stiffness of the bending springs $k^b_i$ is needed. To derive this relation, we consider a classical problem of a cantilever beam length $l_b$, under a point force $\mathbf{F}$ at the free end (figure~\ref{fig:cantilever_beam_point-force}). In the limit of small displacement, the principle of energy yields the value of the bending spring stiffness $k^b$ as a function of the flexural rigidity $EI$. The energy stored in this beam at the static state can be calculated easily using the Euler-Bernoulli beam theory as shown in eqn.~(\ref{eqn:energy_beam}).

\begin{equation}
E_{beam} = \frac{F^2 l_b^3}{6EI} 
\label{eqn:energy_beam}
\end{equation}

The mass-spring network is called an equivalent model of the continuous beam if under the same external loads, its mechanical behavior (in this case, it is the energy stored in the system) is the same as the one of the beam. Let us now study a mass-spring network discretized into $n+2$ mass points connected by bending and extension springs as shown in figure \ref{fig:cantilever_beam_point-force}. All the bending and extension springs are the same with a stiffness $k^b$ and $k^e$, respectively and $k^e \gg k^b$. The first two points are totally fixed to represent the boundary condition of the fixed end of the beam. Writing the equilibrium equations for the remaining $n$ points, we have:

\begin{equation}
F \frac{l_b}{(n+1)} (n+1-i) = k^b (\theta_{i+1} - \theta_{i}) \ \ \rm{for} \ \ i=1...n
\label{eqn:equilibrium_LMS_cantiliver}
\end{equation}

Considering the deformations of extension springs are very small, the total potential energy of all the bending springs of the system is 

\begin{equation}
E_{mass-spring} = \frac{1}{2} k^b \sum_{i=1}^{n} (\theta_{i+1} - \theta_{i})^2
\label{eqn:total_energy_LMS1}
\end{equation}

With eqn. (\ref{eqn:equilibrium_LMS_cantiliver}) and eqn. (\ref{eqn:total_energy_LMS1}) we obtain:

\begin{equation}
\begin{split}
E_{mass-spring} & = \frac{F^2 l_b^2}{2 k^b (n+1)^2} \sum_{i=1}^{n} i^2 \\
                & = \frac{F^2 l_b^2}{2 k^b (n+1)^2} \frac{n(n+1)(2n+1)}{6} \\
                & = \frac{F^2 l_b^2}{12 k^b} \frac{n(2n+1)}{n+1}    
\end{split}
\label{eqn:total_energy_LMS}
\end{equation}

By comparing eqn. (\ref{eqn:energy_beam}) and eqn. (\ref{eqn:total_energy_LMS}), we can derive an analytical relation between $k_b$ and $EI$ as following:

\begin{equation}
k^b = \frac{EI}{l_b} \frac{n(2n+1)}{2(n+1)}
\label{eqn:EI_k_relation}
\end{equation}

Since eqn. (\ref{eqn:EI_k_relation}) is derived based on the assumption of small displacement, we still have here a linear problem thus the principle of superposition can be applied. During the flight, the aerodynamic loads acting on insect wings can be considered to be equivalent to distributed loads on the surface of the wings. These distributed loads can be discretized into many point forces using a work-equivalent method \cite{FEMLogan} and then the superposition principle can be applied. Thus, it is sufficient to analyze only one point force case to find the relation between $EI$ and $k^b$, since it does not depend on the point force $\mathbf{F}$.

However, as mentioned at the beginning of this section, insect wings are deformed significantly to create lift for flying. Here, we are dealing with a large displacement problem and the question is if eqn. (\ref{eqn:EI_k_relation}) still remains valid. The technique used to derive (\ref{eqn:EI_k_relation}) is no longer applicable since the solution for large deflection of a cantilever beam cannot be obtained analytically  \cite{BeamDeflection}. This problem involves calculating elliptical integrals of the second kind \cite{BeamLargeDeflectionSolution} and needs to be solved numerically. Consequently, the relation between $EI$ and $k^b$ is put into a large displacement, nonlinear test case to check if we still get the same mechanical behaviors between the continuous beam and the mass-spring model. The results are presented in the next section.

\subsection{Validations of the mass-spring model}
\label{subsec:validation_MSM}

\subsubsection{Vein model - Cantilever beam under gravitational force}

\paragraph{Static case}
Firstly, we consider a static case of a cantilever beam with length $L=0.3$, flexural rigidity $EI=0.24$ and loaded by a point force $F$ at the free end. The force $F$ varies from $0.39$ to $11.76$ and it must be strong enough to cause a large deflection. All the parameters here are dimensionless. The vertical displacement $\delta y$ and the horizontal displacement $\delta x$ of the free end of the beam at equilibrium state can be calculated by using the fundamental Bernoulli--Euler theorem \cite{BeamLargeDeflection} and the mass-spring network as given in table \ref{tab:comparison_EI_to_k}.The static state of the vein model (discretized by $n = 64$ mass points) is obtained by solving the dynamic equations of the system with artificial damping forces to make the system go quickly towards its balanced position. Despite the small displacement assumption for deriving the relation between $EI$ and $k^b$, the relation in eqn. (\ref{eqn:EI_k_relation}) is still valid even in very large deflection problem. For the case $F=3.92$, the vertical displacement of the free end is already more than $30 \%$ of the total length of the beam and we still get very good agreements between both models with the relative error being smaller than $1 \%$. The mapping from $EI$ to $k^b$ can then be generalized for nonlinear, large deflection cases with good agreement between the continuum theory and the discrete mass-spring network.

\begin{table*}[tb]
\caption{Comparison between the continuum theory and the discrete mass-spring network in the static large deflection case.}
 \centering\small
 \begin{tabular}{c c c c c c c} 
 \toprule
 Point force & \multicolumn{2}{c}{Nonlinear beam \cite{BeamLargeDeflectionSolution}} & \multicolumn{2}{c}{Mass-spring network} & \multicolumn{2}{c}{Relative error} \\ [0.5ex] 
 $F$ & $\delta x_{ref} [10^{-2}]$ & $\delta y_{ref} [10^{-2}]$ & $\delta x [10^{-2}]$ & $\delta y [10^{-2}]$ & $err_x [\%]$  & $err_y [\%]$  \\ [0.5ex]
 \hline\hline \\ [-2ex]
 0.39 & 29.96 & -1.46 & 29.96 & -1.46 & 0 & 0  \\ [0.5ex] 
 \hline \\ [-2ex]
 1.96 & 29.02 & -6.93 & 29.01 & -6.92 & 0.03 & 0.14 \\ [0.5ex] 
 \hline \\ [-2ex]
 3.92 & 26.87 & -12.14 & 26.85 & -12.1 & 0.07 & 0.33 \\ [0.5ex] 
 \hline \\ [-2ex]
 7.84 & 22.53 & -17.93 & 22.53 & -17.85 & 0 & 0.45 \\ [0.5ex] 
 \hline \\ [-2ex]
 11.76 & 19.37 & -20.69 & 19.40 & -20.57 & 0.15 & 0.58 \\ [0.5ex]
 \bottomrule
\end{tabular}

 \label{tab:comparison_EI_to_k}
\end{table*}

\paragraph{Dynamic case}

\begin{figure}[ht]
\centering
\includegraphics[width=0.45\textwidth]{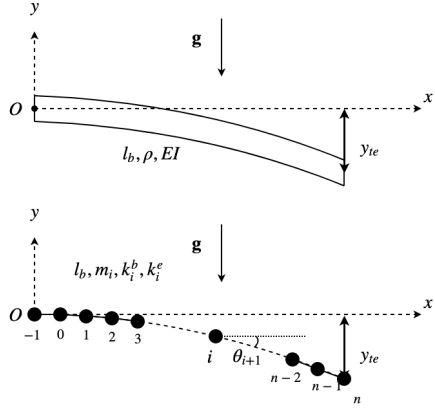}
\caption{Cantilever beam subjected to gravity field modeled by continuous nonlinear beam and mass-spring network.}
\label{fig:cantilever_beam_gravity}
\end{figure}

 The vein model will now be compared with another solid solver developed by Engels et al.~\cite{ThomasElasSwimmer}. It is based on the classical nonlinear beam equation, the Euler--Bernoulli theory. All details about this solver can be found in \cite{ThomasThesis}. We study the case when we have a 2D cantilever beam (figure \ref{fig:cantilever_beam_gravity}) of length $l_b=1$, density $\rho=0.0571$, flexural rigidity $EI=0.0259$. The beam is in vacuum, subjected to a gravity field $g=0.7$ strong enough to cause large deflection. All the parameters here are dimensionless. Both computations are performed for the same numerical parameters with the time step $dt=10^{-2}$ and $n=64$ grid points. Although both solvers require the same amount of CPU time for the same resolution, the mass-spring network is still more efficient since it is designed to deal with 3D problems. For the computation, the mass-spring solver handles a system of $3 n$ degrees of freedom, corresponding to 3 dimensions, while the nonlinear-beam solver only solves for $2 n$ variables.
 
\begin{figure}[h]
\centering
\begin{minipage}{.5\linewidth}
\centering
  \includegraphics[width=1.075\linewidth]{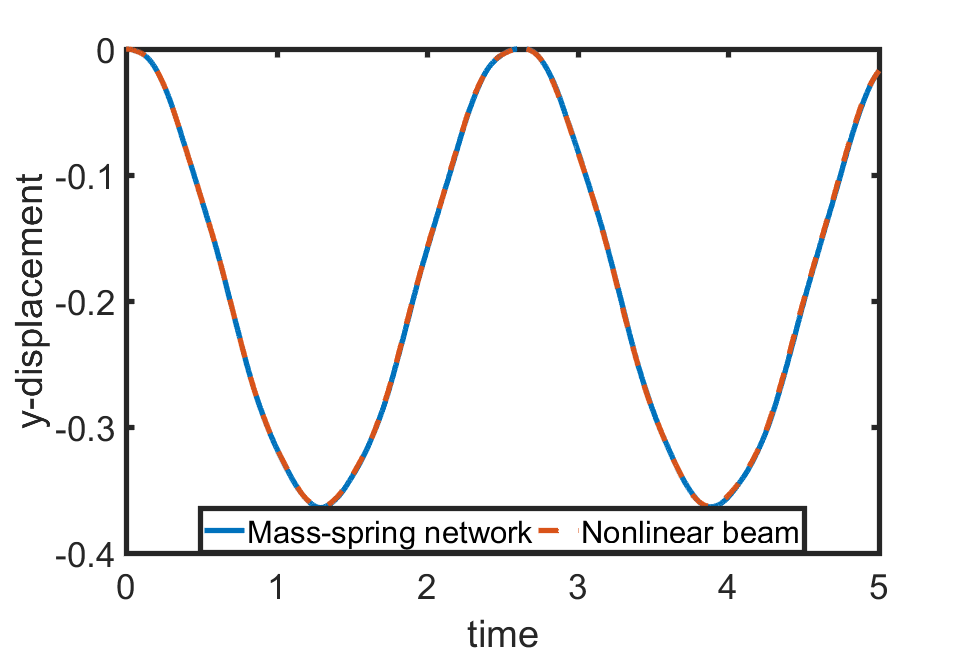}
  \caption*{(a)}
\end{minipage}%
\begin{minipage}{.5\linewidth}
\centering
  \includegraphics[width=1.075\linewidth]{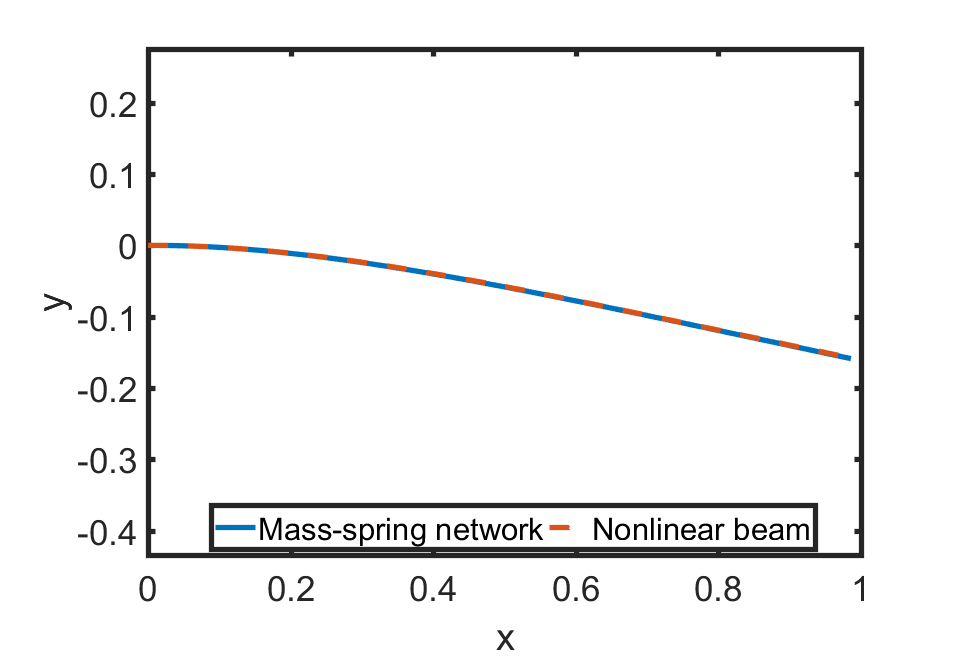}
  \caption*{(b)}
\end{minipage}

\caption{The oscillation of the trailing edge $y_{te}(t)$ (a) and the deflection lines at $t=2$ (b) calculated by the nonlinear beam theory \cite{ThomasThesis} (dashed blue line) and the mass-spring network (solid red line).}
\label{fig:validation_nonlinear_beam}
\end{figure}

 The deflection line of the two models at a given time $t$ and the oscillation of the trailing edge $y_{te}(t)$ are shown in figure~\ref{fig:validation_nonlinear_beam}. The dashed blue line calculated by the nonlinear beam theory and the solid red line calculated by the mass-spring network are almost coincident with each other. We have an excellent agreement between these two models with a relative error smaller than $1\%$. 

\subsubsection{Membrane model - Uniaxial and isotropic deformations of a two-dimensional sheet}

We consider here the same test case proposed by Omori et al.~\cite{MembraneOmori} where a square 2D sheet with an initial side length $l_0=1$ is extended by a uniaxial tension $T=0.005$ and has a final length $l$ in the $x$-direction at the equilibrium state, as shown in figure \ref{fig:mesh_type_for_membrane}. This tension must be small enough to cause small deformation on the sheet. The Young modulus $E_s$ is defined by:

\begin{equation}
E_s = \frac{T}{\epsilon}
\label{eqn:Young_modulus_definition}
\end{equation}

where $\epsilon$ is the strain.

\begin{figure}[th]
\centering
\includegraphics[width=0.45\textwidth]{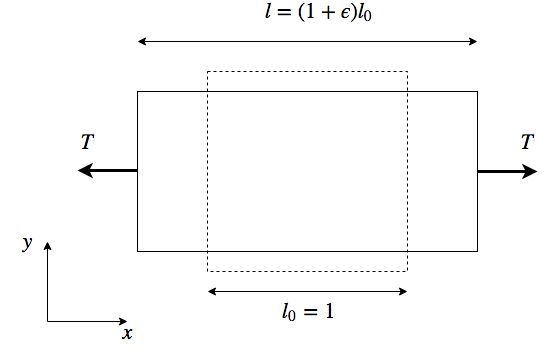}
\caption{Deformation of a 2D sheet along the $x$-axis under the uniaxial tension $T$.}
\label{fig:2D_sheet_uniaxial_extension}
\end{figure}

\begin{figure}[ht]
\centering
\includegraphics[width=0.45\textwidth]{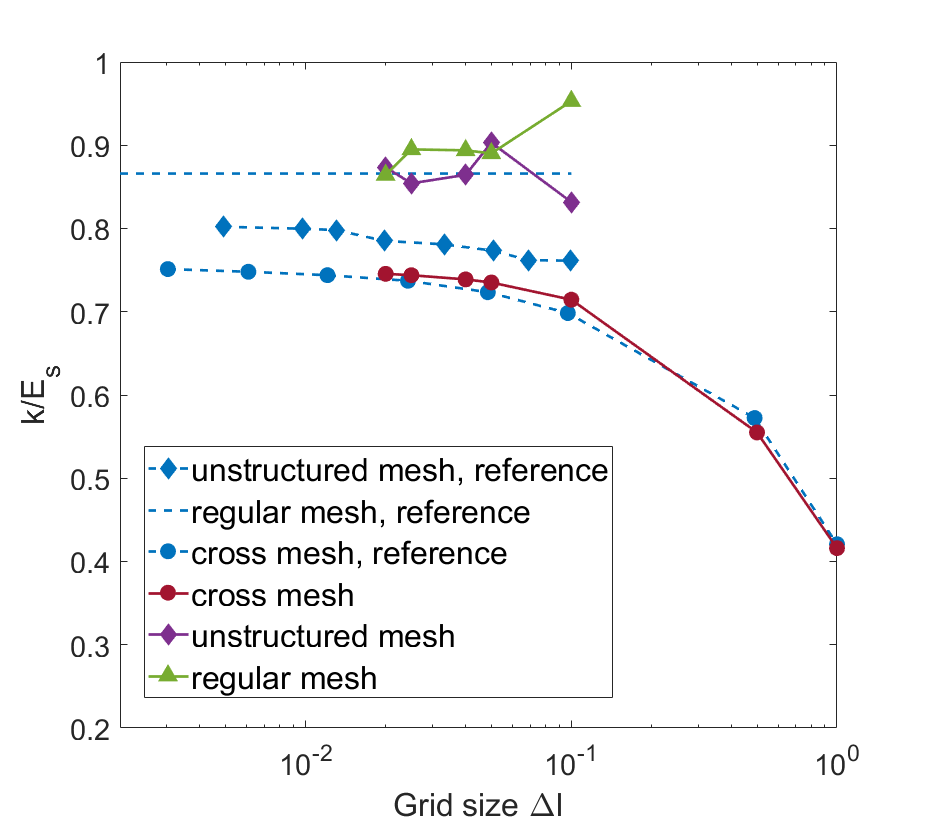}
\caption{Effect of mesh topology on the relation between the spring stiffness $k$ and the corresponding Young modulus $E_s$}
\label{fig:Mesh_dependency}
\end{figure}

The sheet is then discretized by using three types of meshes, illustrated in figure~\ref{fig:mesh_type_for_membrane}. The grid size $\Delta l$ is the side length of one triangle element of the mesh and inversely proportional to the number of grid points $n$. The grid size is varied to refine the mesh resolution. Since we are only interested in the equilibrium state of the model, the masses will not have any effect on the result and they are chosen properly for the numerical convergence. Instead of solving the static equation of the system, we still solve here the dynamic equations of the system with artificial damping forces to make the system go quickly towards its balanced position. Last but not least, all the spring stiffnesses are set to the same value $k=1$. Figure \ref{fig:Mesh_dependency} shows the results we get for all three mesh structures. First, for the cross-center structure, we are able to reproduce the result of Omori et al. \cite{MembraneOmori}. When the mesh is refined, the ratio $k/E_s$ converges to the analytical value $3/4$ with the relative error being smaller than $1\%$. For the regular triangle, due to the shape of the square, we have some minor flaws of the mesh at the border. But these can be neglected when the mesh is fine enough and we can consider it as a regular triangular mesh. Indeed, for high resolution, we find again an excellent agreement with the analytical ratio $k/E_s=\sqrt{3}/2$ derived by Omori et al. The relative error is also smaller than $1\%$. However, for the unstructured mesh, the convergent value of $k/E_s$ is larger than the one of Omori et al., but identical to the analytical solution for the regular triangle. This finding is in fact expected by Omori et al. since these two meshes are both constructed with triangles, each node being connected to six springs. Yet, the random structure of the unstructured mesh makes it difficult to explain the difference.

Using our mass-spring model, we are capable of reproducing the results from the references, which indicates the reliability of the solver. These results for both small and large deformation cases allow us to have the same conclusions as in the literature about the mass-spring system. Even though the mechanical properties of spring networks are strongly dependent on the mesh topology, a correlation between the discrete model and the continuum model can still be obtained if the mesh is fine enough. However, this needs to be compromised with the computational efficiency which is the main reason that we choose mass-spring network in the first place.

\section{Wing structure} \label{sec:Wing_structure}

The simulation of insects with flexible wings is extremely complicated not only because it involves solving for both fluid and solid dynamics, but also due to the fact that insect wings are sophisticated structures. In our work, we want to take into account as much as possible all the mechanical properties of the bumblebee wing, in order to correctly model its dynamic behaviors. In our wing model there are three main factors introduced, which are considered to have the most impact on wing deformation during flight: venation pattern, mass distribution and vein stiffness.   

\subsection{Venation pattern}

The venation architecture is claimed to affect the anisotropy of the wing. Throughout measurements from different insects, Combes et al. \cite{CombesI,CombesII} suggest that wing flexural stiffness declines exponentially towards the tip and trailing edge. This is explained by the common venation patterns of insect wings: most insect wings have thick, stiff veins at the leading edge and cross veins are thinner as they expand toward the wing tip. This structure allows insect wings to resist against strong bending deformation in the spanwise direction, while creating camber for lift generation in the chordwise direction \cite{CombesII}. Nakata and Liu~\cite{FSINakata2012} modeled the anisotropy caused by wing veins. To this end they took into account the variation of wing thickness and introduced a "rule of mixture" of composite materials.

\begin{figure}[ht]
\centering
\includegraphics[width=.9\linewidth]{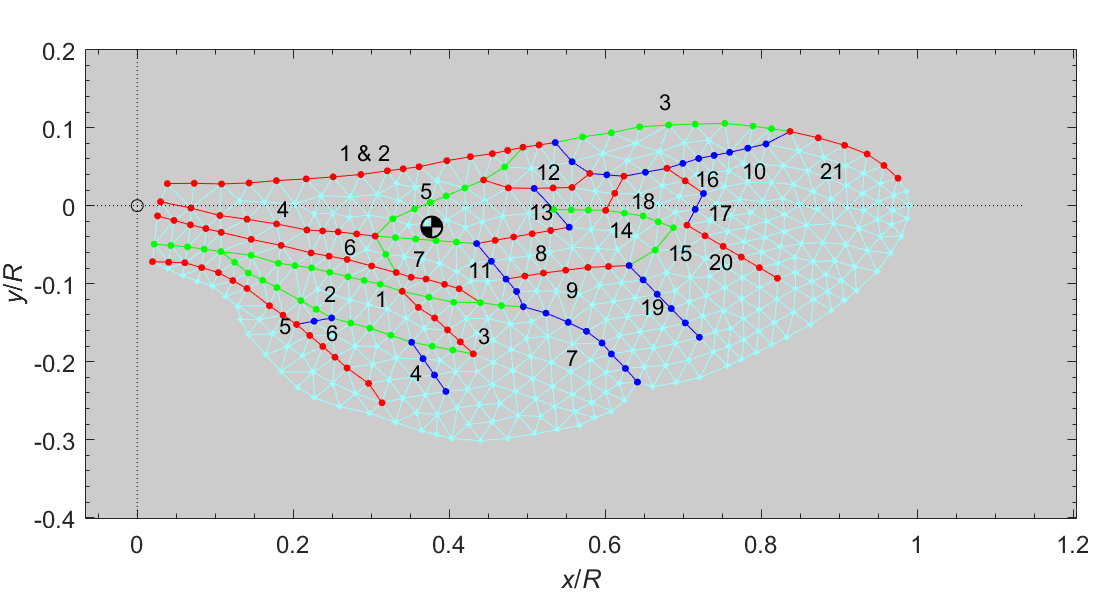}\\
  \caption{Illustration of the mass-spring model which is meshed based on measured data of real bumblebee wings \cite{BumblebeeWingStructure}. The black and white markers represent mass centers. Color codes (red, green and blue) are used for identifying veins and the membranes are represented by cyan circles.}
  \label{fig:wing_mesh}
\end{figure}

In our model, the functional approach is used to take into consideration the venation pattern. The vein structure, as well as the wing contour, are adapted from the data from \cite{BumblebeeWingStructure} and encoded into the mass-spring network, as shown in figure~\ref{fig:wing_mesh}. Comparing to the reference data, two more veins are added (vein 21 in the forewing and vein 7 in the hindwing) and two forewing veins 19 and 20 are extended toward the tip of the wing. These modifications are made to add bending stiffness to the tip of the wing and thus to obtain a more realistic behavior during the simulation. The meshing is done by identifying firstly the contour of the wing and all the veins (green, red and blue curves in figure~\ref{fig:wing_mesh}). The membrane is then discretized by a triangular mesh using SALOME \footnote{https://www.salome-platform.org/}, an open-source integration platform for numerical simulation and mesh generation. 

\subsection{Mass distribution}

Another property which plays an essential role on the aerodynamics of the wing, is the mass distribution. It represents the inertia of the system and the position of the mass center has a strong connection with the stability of the wing during flight. The mass distribution is calculated based on the measured wing mass data from \cite{BumblebeeWingStructure}. For our numerical simulations, the total wing mass is chosen as the same used by Kolomenskiy et al.~\cite{BumblebeeWingStructure}, $m_w = 0.791~mg$. The mass is then distributed into vein and membrane parts based on their geometry and material. Each vein is considered as a rod composed of cuticle, $\rho_c = 1300~kg/m^3$ \cite{BumblebeeWingStructure}, with a circular cross section of constant diameter $d_v$ \cite{BumblebeeWingStructure} and length $l_v$, calculated directly from the model. The mass of each vein is then calculated and shown in table~\ref{table:vein_mass}. Both diameter and mass are dimensionless quantities, normalized by wing length $L$ and air density $\rho_{air} L^3$, respectively.

\begin{table}[]
\centering
\begin{tabular}{|c|c|c|ccc}
\hline
\multicolumn{3}{|c|}{Forewing}                                                                                           & \multicolumn{3}{c|}{Hindwing}                                                                                                                                                           \\ \hline
\# & \begin{tabular}[c]{@{}c@{}}Nominal\\ diameter\end{tabular} & \begin{tabular}[c]{@{}c@{}}Nominal\\ mass\end{tabular} & \multicolumn{1}{c|}{\#} & \multicolumn{1}{c|}{\begin{tabular}[c]{@{}c@{}}Nominal\\ diameter\end{tabular}} & \multicolumn{1}{c|}{\begin{tabular}[c]{@{}c@{}}Nominal\\ mass\end{tabular}} \\ \hline
1  & 0.0070                                                     & 0.0209                                                 & \multicolumn{1}{c|}{1}  & \multicolumn{1}{c|}{0.0065}                                                     & \multicolumn{1}{c|}{0.0180}                                                 \\ \hline
2  & 0.0074                                                     & 0.0237                                                 & \multicolumn{1}{c|}{2}  & \multicolumn{1}{c|}{0.0043}                                                     & \multicolumn{1}{c|}{0.0071}                                                 \\ \hline
3  & 0.0055                                                     & 0.0076                                                 & \multicolumn{1}{c|}{3}  & \multicolumn{1}{c|}{0.0046}                                                     & \multicolumn{1}{c|}{0.0024}                                                 \\ \hline
4  & 0.0070                                                     & 0.0063                                                 & \multicolumn{1}{c|}{4}  & \multicolumn{1}{c|}{0.0011}                                                     & \multicolumn{1}{c|}{0.0001}                                                 \\ \hline
5  & 0.0040                                                     & 0.0031                                                 & \multicolumn{1}{c|}{5}  & \multicolumn{1}{c|}{0.0038}                                                     & \multicolumn{1}{c|}{0.0043}                                                 \\ \hline
6  & 0.0048                                                     & 0.0094                                                 & \multicolumn{1}{c|}{6}  & \multicolumn{1}{c|}{0.0037}                                                     & \multicolumn{1}{c|}{0.0005}                                                 \\ \hline
7  & 0.0040                                                     & 0.0019                                                 & \multicolumn{1}{c|}{7}  & \multicolumn{1}{c|}{0.0020}                                                     & \multicolumn{1}{c|}{0.0012}                                                 \\ \hline
8  & 0.0038                                                     & 0.0009                                                 &                         &                                                                                 &                                                                             \\ \cline{1-3}
9  & 0.0041                                                     & 0.0023                                                 &                         &                                                                                 &                                                                             \\ \cline{1-3}
10 & 0.0048                                                     & 0.0064                                                 &                         &                                                                                 &                                                                             \\ \cline{1-3}
11 & 0.0045                                                     & 0.0017                                                 &                         &                                                                                 &                                                                             \\ \cline{1-3}
12 & 0.0038                                                     & 0.0018                                                 &                         &                                                                                 &                                                                             \\ \cline{1-3}
13 & 0.0042                                                     & 0.0010                                                 &                         &                                                                                 &                                                                             \\ \cline{1-3}
14 & 0.0038                                                     & 0.0020                                                 &                         &                                                                                 &                                                                             \\ \cline{1-3}
15 & 0.0034                                                     & 0.0008                                                 &                         &                                                                                 &                                                                             \\ \cline{1-3}
16 & 0.0032                                                     & 0.0005                                                 &                         &                                                                                 &                                                                             \\ \cline{1-3}
17 & 0.0032                                                     & 0.0004                                                 &                         &                                                                                 &                                                                             \\ \cline{1-3}
18 & 0.0044                                                     & 0.0009                                                 &                         &                                                                                 &                                                                             \\ \cline{1-3}
19 & 0.0015                                                     & 0.0001                                                 &                         &                                                                                 &                                                                             \\ \cline{1-3}
20 & 0.0018                                                     & 0.0001                                                 &                         &                                                                                 &                                                                             \\ \cline{1-3}
21 & 0.0020                                                     & 0.0009                                                 &                         &                                                                                 &                                                                             \\ \cline{1-3}
\end{tabular}

\caption{Dimensionless vein diameter $d_v$ (adapted from \cite{BumblebeeWingStructure}) and their corresponding dimensionless mass $m_v$.}
\label{table:vein_mass}
\end{table}

The mass distribution for the membrane is more tricky since we do not have the material properties of bumblebee membranes. A bi-linear regression is employed due to the fact that the membrane is heavier near the wing root and the leading edge \cite{BumblebeeWingStructure}. An optimization is employed to find the parameters for the regression using the mass center from the measured data as an objective function. For a mass point $m_i$ belonging to the membrane at position $[x_i, y_i]$ (the $z$ coordinate is neglected here because we assume that the membrane is a planar sheet), we get:
\begin{equation}
    m_i = 1.75 \cdot 10^{-4} - 2.83 \cdot 10^{-4} x_i + 4.91 \cdot 10^{-4} y_i
\end{equation}
This yields a difference, between two mass centers, of $0.0013$ in the $x$-direction and $0.0085$ in the $y$-direction which are really small compared to the reference wing length $R_w = 1$.

\subsection{Vein stiffness estimation}

To study the influence of wing flexibility on the aerodynamics performance, the flexural rigidity of veins will be changed to alter the bending stiffness of the wing.  Consequently, only Young modulus $E$ will be varied. Insect cuticles are reported to have a Young modulus about $1 kPa$ to $50 MPa$ \cite{CuticleProperties}. For our study, we are choosing two values of $E$ (corresponding to flexible and highly flexible wings): $3.5~kPa$ and $35~kPa$, which are in this range and give realistic deformations comparing to those observed in real life. Then, the flexural rigidity $EI$ of each vein will be calculated using the second moment of inertia $I$ of circular-section veins with diameters given in table~\ref{table:vein_mass}.

\section{Fluid-structure interaction}\label{sec:FSI}

\subsection{Numerical method}

The ultimate goal of this work is the fluid-structure interaction simulation of insects with flexible wings. To study the airflow as well as the effect of flexibility on the aerodynamic performance of the wing, the developed mass-spring model needs to be coupled with a fluid solver. This is done by combining the volume penalization method \cite{VolPenaAngot} with a Fourier pseudospectral discretization \cite{VolPenaDmitry}, for which we developed the parallel open–source solver FLUSI, freely available on Github\footnote{https://github.com/pseudospectators/FLUSI} \cite{Flusi}. The code solves the incompressible, penalized Navier-–Stokes equations

\begin{align}
\partial_t \mathbf{u} + \mathbf{\omega} \times \mathbf{u} &= - \nabla \Pi + \nu \nabla^2 \mathbf{u} - \underbrace{\frac{\chi}{C_{\eta}} (\mathbf{u} - \mathbf{u}_s)}_{\text{penalization term}} - \underbrace{\frac{1}{C_{sp}} \nabla \times \frac{(\chi_{sp} \mathbf{\omega})}{\nabla^2}}_{\text{sponge term}} \\
\nabla \cdot \mathbf{u} &= 0 \\
\mathbf{u} (\mathbf{x}, t=0) &= \mathbf{u}_0 (\mathbf{x}) \ \ \ \ \ \ \mathbf{x} \in \Omega, t>0
\end{align}
where $\mathbf{u}$ is the fluid velocity, $\mathbf{\omega}$ is the vorticity, $\Pi = p +\frac{1}{2}\mathbf{u} \cdot \mathbf{u}$ is the total pressure, $\nu$ is the kinematic viscosity. We find here again all the familiar terms of the classical Navier--Stokes equations except for the sponge and penalization terms. The former is a vorticity damping term used to gradually damp vortices and alleviate the periodicity inherent to the Fourier discretization. The last term is used to impose the no-slip boundary conditions on the fluid-solid interface as explained in \cite{Flusi}. All the geometry information of the solid is encoded in the mask function $\chi$, which is usually taken as one inside the solid and zero otherwise. However, we are dealing with a moving flexible obstacle and the discontinuous mask function $\chi$ need to be replaced by a smooth one, eqn. (\ref{eqn:mask_function}), to avoid oscillations in the hydrodynamic forces \cite{ThomasThesis}. Thus, we employ a mask function $\chi$ as shown below:

\begin{equation}
    \chi(\delta)= 
\begin{cases}
    1 &  \delta \leq t_w-h\\
    \frac{1}{2} \left(1+\cos{\pi \frac{(\delta-t_w+h)}{2h}}\right) & t_w-h < \delta < t_w+h \\
    0 & \delta \geq t_w+h
\end{cases}
\label{eqn:mask_function}
\end{equation}
where $h$ is the semi-thickness of the smoothing layer, $t_w$ is the semi-thickness of the wing and $\delta$ is the distance function which gives us the distance from Eulerian fluid nodes to the center surface of the wing. As presented in section \ref{sec:Wing_structure}, an unstructured triangular mesh is employed for our wing model. Thus, the discretized wing surface is composed of triangles constructed by three vertices (e.g. $\mathbf{x}_{s,i}, \mathbf{x}_{s,j}$ and $\mathbf{x}_{s,k}$). The distance function $\delta$ is computed by cycling over all these triangles. Since we are only interested in the fluid nodes near the fluid-solid interface, a bounding box is used to save computing time. For each triangle, the distance from it to all the fluid nodes belonging to its bounding box will be computed by using the algorithm from \cite{TriDistance}. The distance function at one fluid node is finally the minimum distance from this fluid node to all the triangles nearby. 

\begin{equation}
    \delta(\mathbf{x},t) = \min (|| \mathbf{x} - triangle(\mathbf{x}_{s,i}, \mathbf{x}_{s,j}, \mathbf{x}_{s,k}) ||_2)
\end{equation}
The solid velocity field $\mathbf{u}_s$ is calculated in the same way as the distance function $\delta$. If the triangle $(\mathbf{x}_{s,i}, \mathbf{x}_{s,j}, \mathbf{x}_{s,k})$ is the one closest to the fluid node $\mathbf{x}$, $\mathbf{x}$ will be projected onto the plane of the triangle and the solid velocity of the projected point is interpolated from the velocities of the three vertices by using barycentric interpolation. Because we do not consider the flexibility of the wing in the direction perpendicular to the wing surface, the velocity of the projected point should be the same as the one of the fluid node. Nevertheless, the solid velocity field is defined in the global reference frame for the fluid solver while the velocity solved by the solid solver is in the local wing reference frame. These velocities are needed to be transformed back to the global reference frame using eqn. (\ref{eqn:relative_velocities}) where $\mathbf{V}_{O^{(w)}}$ and $\mathbf{\Omega}$ are the translating and rotating velocity of the wing reference frame, $\mathbf{v}^{(w)}$ and $\mathbf{x}^{(w)}$ are the velocity and the position computed by the solid solver in the wing reference frame, respectively.

\begin{equation}
    \mathbf{u}_s = \mathbf{V}_{O^{(w)}} + \mathbf{v}^{(w)} + \mathbf{\Omega} \times \mathbf{x}^{(w)}
\label{eqn:relative_velocities}
\end{equation}

Moreover, the fluid also interacts with the wing via the pressure and viscous force. However, at $Re = \mathcal{O}(10^{3})$ in our study, the viscous force is considered to be very small and only the pressure force is transferred into the mass-spring system as external force. Contrary to the previous calculation of the solid velocity field $\mathbf{u}_s$, the pressure force is interpolated from the Eulerian fluid grid onto the Lagrangian mass points. The pressure interpolation is quite straightforward because the pressure field solved by the penalized Navier--Stokes equations is following the Darcy law and continuous even inside the wing \cite{VolPenaAngot}. Consequently, the pressure at any mass point can always be determined, using delta interpolation proposed by Yang et al.~\cite{DeltaInterp}, from pressure values at neighboring Eulerian grid points~\cite{ThomasThesis}. Then for each triangle element of the wing, the pressure forces at the three vertices are perpendicular to the triangle and have magnitudes equaling to the pressure multiplied by one third of the triangle area. This is done for all the triangles and then accumulated to obtain the overall external pressure forces acting on the mass-spring system.

For time-stepping, the coupled fluid-solid system is advanced by employing a semi-implicit staggered scheme as proposed in \cite{ThomasThesis}. At time step $t^n$, the fluid velocity field is advanced to new time level $\mathbf{u}^n \rightarrow \mathbf{u}^{n+1}$ from the old mask function $\chi^n$ and the old solid velocity field $\mathbf{u}^n_s$ by using the Adams--Bashforth scheme. Then, the pressure field at the new time step $p^{n+1}$ is calculated from the fluid velocity field $\mathbf{u}^{n+1}$. Finally, the solid is advanced to the new time step $\chi^{n+1}$ and $\mathbf{u}^{n+1}_s$ and the whole process is repeated until the final time.

\subsection{Validation}

 For the validation of the fluid-solid coupling, we consider two test cases: the Turek benchmark test case FSI3 \cite{Turek1,Turek2} and the rigid revolving bumblebee wing test case \cite{RigidRevolWing}.

\subsubsection{Turek benchmark FSI3}

\begin{figure}[th]
\centering
\includegraphics[width=0.45\textwidth]{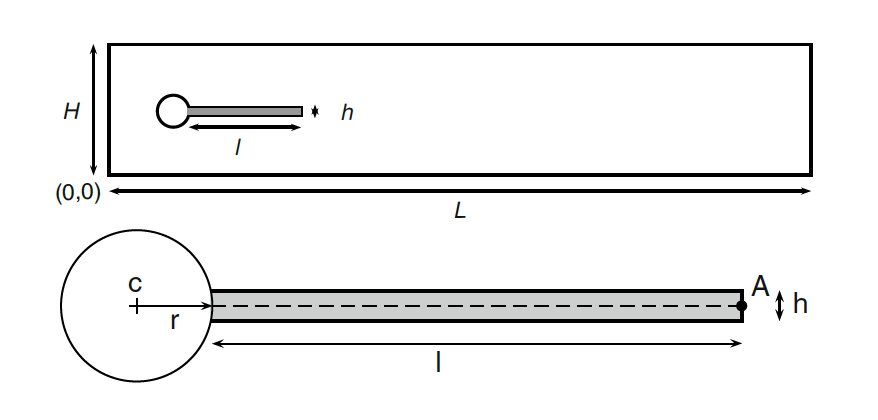}
\caption{Computational domain of the FSI3 Turek benchmark and dimensions of the solid part \cite{Turek2}.}
\label{fig:Turek_benchmark_config}
\end{figure}

 The Turek benchmark FSI3 involves a flexible appendage of length $l=0.35$ and thickness $h=0.02$ placed right behind a circle cylinder of radius $R=0.05$; the whole obstacle is immersed in a channel of size $H \times L = 0.41 \times 2.5$ with a Poiseuille inflow of meanflow $\bar{U}=2$, as shown in figure \ref{fig:Turek_benchmark_config}. The center of the cylinder is placed a bit lower to the centerline at $(0.2, 0.2)$ to trigger the instability and to make the appendage oscillate.
 
 The fluid solver, as well as the fluid-solid coupling, are handled by the software FLUSI \cite{Flusi} and the setup remains the same as the test case FSI3 with a resolution of $5200 \times 1152$ whose details can be found in \cite{ThomasThesis}. Only the solid solver based on the nonlinear beam equation, which is used for validation in section \ref{sec:flex}, is now replaced by the new solver using the mass-spring network for validation. The results of this simulation are presented in table~\ref{tab:Turek_benchmark_validation} for the comparison with the reference solutions presented in the literature \cite{ThomasThesis,Turek1,Turek2}.

\begin{figure}
\centering
  \includegraphics[height=.45\linewidth]{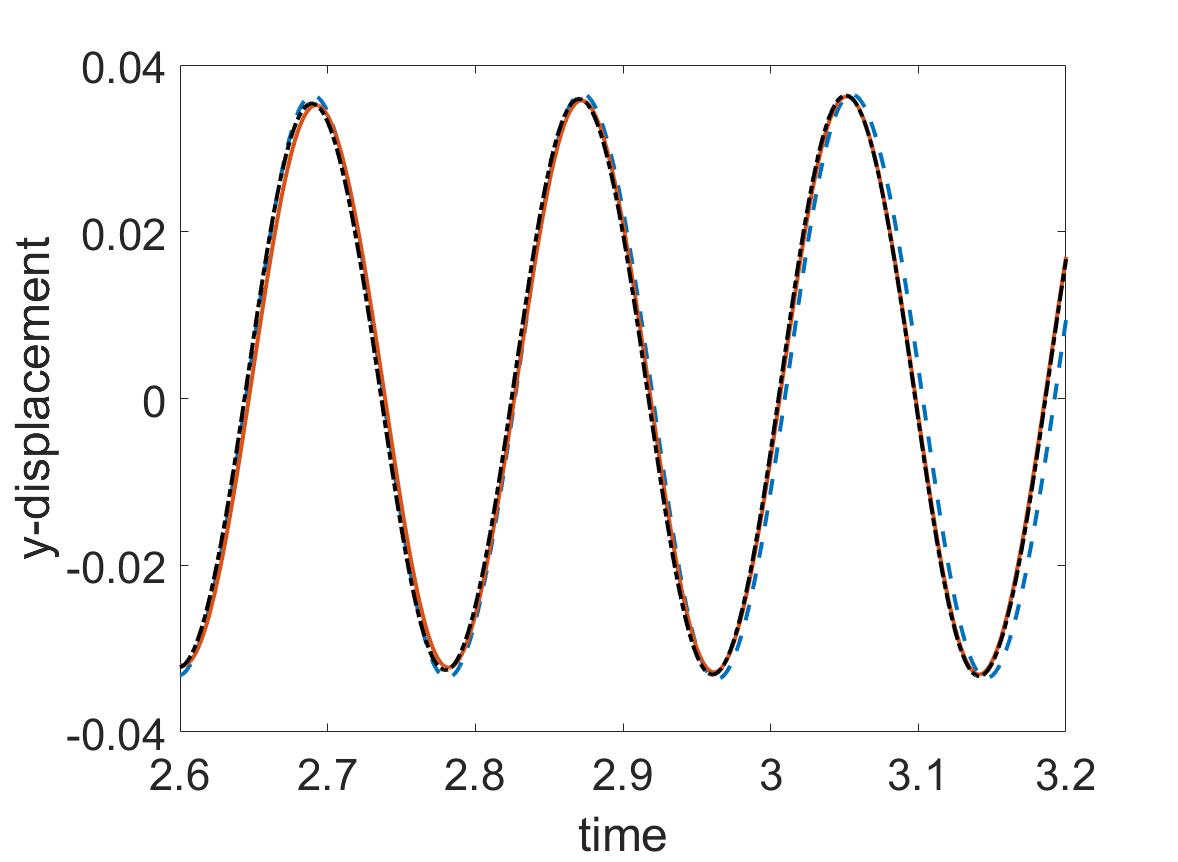}
  \caption*{(a)}
\par\medskip
\begin{minipage}{.5\linewidth}
\centering
\includegraphics[height=.72\linewidth]{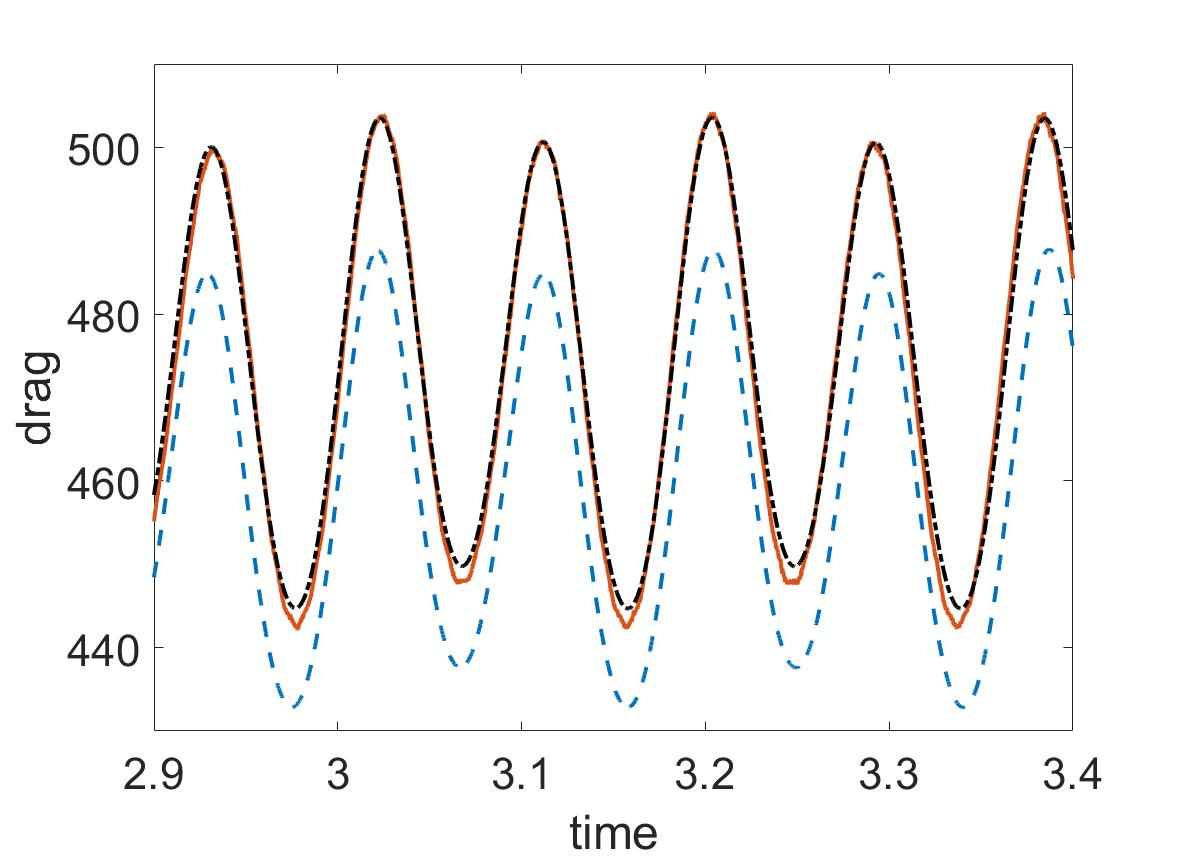}
\caption*{(b)}
\end{minipage}%
\begin{minipage}{.5\linewidth}
\centering
\includegraphics[height=.72\linewidth]{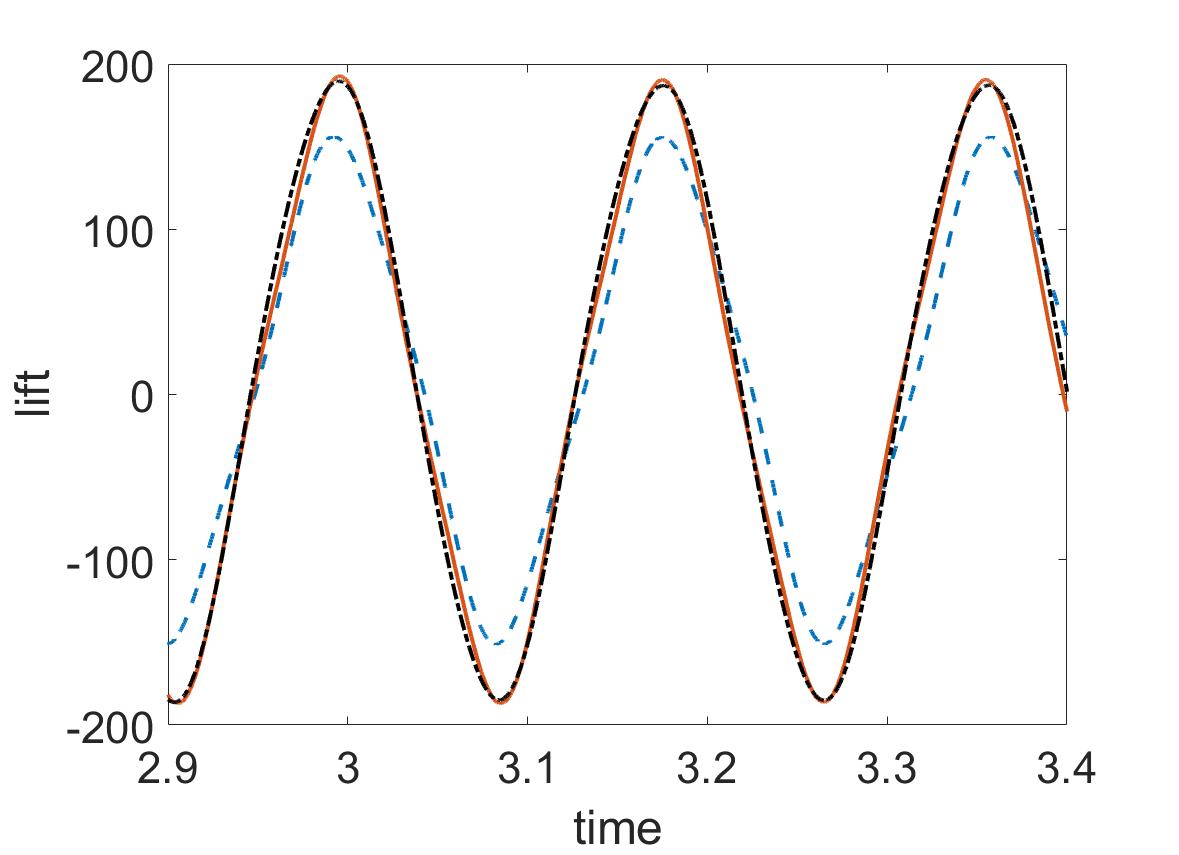}
\caption*{(c)}
\end{minipage}

\caption{The oscillation of the trailing edge $y_{te}(t)$ (a), the drag (b) and the lift (c) from the references \cite{Turek2} (blue dashed line), \cite{ThomasThesis} (black dash-dotted line) and the mass-spring network (red continuous line).}
\label{fig:Turek_benchmark}
\end{figure}

For the oscillation of the trailing edge $y_{te}$, the result is in excellent agreement with all three references when the maximum relative error, for both maximum and minimum values of $y_{te}$, is only $1.76 \%$ and the relative error for the frequency of oscillation is $1.65 \%$. The vertical displacement of the trailing edge with respect to time in the periodic state is also plotted in figure~\ref{fig:Turek_benchmark} to compare with the reference \cite{Turek2}. The two lines are almost superposed on each other. Nevertheless, the computed drag is less accurate with a relative error which can go up to $4.57 \%$ comparing the maximum value with the reference \cite{ThomasThesis}, but only $1 \%$ comparing with \cite{Turek1}. From figure \ref{fig:Turek_benchmark}, the curves of the two solutions appear to have the same shape but have some offset. This offset is explained in \cite{ThomasThesis} to be due to the smoothing layer in the mask function, which plays a role as surface roughness. This leads to the over-prediction of the drag force. Concerning the lift force, the mass-spring model yields results very close to the one calculated by Engels with the error of $2.76\%$, and the difference is around $20\%$ with respect to \cite{Turek1,Turek2} for both max and min values. Like in Engels~\cite{ThomasThesis}, the amplitude of the lift force is over-predicted by coupling FLUSI and the mass-spring solver. In conclusion we find satisfying agreement with the results from the literature, for the solid solver alone, as well as for the FSI algorithm coupling the solid solver with the fluid solver in 2D.

\begin{table*}
\centering
 \begin{tabular}{c c c c c c c c} 
 \hline
 References & \multicolumn{2}{c}{$y_{te} [10^{-3}]$} & \multicolumn{2}{c}{Drag} & \multicolumn{2}{c}{Lift} & $f_0$ \\ [0.5ex] 
 \ \ & max & min & max & min & max & min & \ \ \\ [0.5ex] 
 \hline\hline \\ [-1.5ex]
 Mass-spring network & 36.22 & -32.93 & 503.02 & 442.12 & 189.94 & -186.23 & 5.56  \\ [0.5ex] 
 \hline\hline \\ [-1.5ex]
 (1) T. Engels \cite{ThomasThesis} & 35.63 & -32.71 & 481.20 & 432.50 & 188.52 & -181.30 & 5.44 \\ [0.5ex] 
 \hline \\ [-1.5ex]
 (2) S. Turek \cite{Turek1} & 36.37 & -33.45 & 487.81 & 432.79 & 156.13 & -151.31 & 5.47 \\ [0.5ex] 
 \hline \\ [-1.5ex]
 (3) S. Turek \cite{Turek2} & 36.46 & -33.52 & 488.24 & 432.76 & 156.40 & -151.40 & 5.47 \\ [0.5ex] 
 \hline
\end{tabular}
\caption{Results of the FSI3 benchmark.}
 \label{tab:Turek_benchmark_validation}
\end{table*}

\subsubsection{Rigid revolving wing}

\begin{figure}[ht]
\centering
  \begin{tabular}{@{}c@{}}
     \includegraphics[width=0.45\textwidth]{pictures/rotating_wing_configuration.PNG}\\
    \small (a) Scheme of the revolving wing, adapted from~\cite{RigidRevolWing}.
  \end{tabular}

  \begin{tabular}{@{}c@{}}
    \includegraphics[width=0.45\textwidth]{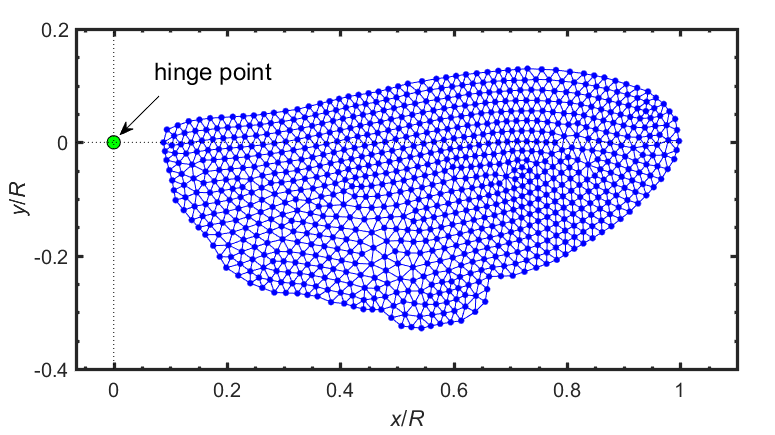}\\
    \small (b) Corresponding discrete mass-spring network.
  \end{tabular}
  \caption{Schematic diagram of the revolving wing simulation (a) and the wing mass-spring model (b) adapted from~\cite{RigidRevolWing}. The wing is rotated around the hinge point with an angle of attack $\alpha = 45^{\circ}$.}
  \label{fig:revolving_wing_config}
\end{figure}

Prior to studying the flexibility of the wing, a common test case of a rigid revolving wing is considered to validate the coupling between the fluid and the solid solver in 3D, i.e. the mask function generation and the velocity field of the solid. The setup is taken the same as the one used by Engels et al.~\cite{RigidRevolWing} as shown in figure~\ref{fig:revolving_wing_config}. 
The angle of attack is fixed at $\alpha = 45^\circ$ while the rotation angle $\phi(t)$ is given by

\begin{equation}
    \phi(t) = \tau e^{{-t}/{\tau}} + t - \tau
\label{eqn:revolving_angle}
\end{equation}

The wing is rotated around the center of the computational domain of size $4 \times 4 \times 2$, which is discretized using a mesh of $1024 \times 1024 \times 512$ grid points. To be consistent with the reference simulation~\cite{RigidRevolWing}, the wing shape is not the one presented in section~\ref{sec:Wing_structure} but adapted from the wing planform taken from the reference. The wing shape is then discretized by a triangular mesh as shown in figure~\ref{fig:revolving_wing_config}b. However, the vein structure will not be taken into account in this model because we are considering a rigid wing. The triangular mesh is solely exploited for the creation of the mask function and the solid velocity field by using the algorithm presented at the beginning of this section. Here, all the quantities are normalized. The wing length is chosen as the length scale $L  = 13.2~mm$; the mass scale is based on the air density $M = \rho_{air} \times L^3 = 2.817~mg$ and the time scale is chosen in the way that wing tip velocity is unity, thus $T = 1~s$ . The Reynolds number is then defined as in \cite{RigidRevolWing} $Re = \bar{u}_{tip} c_m / \nu$ where the mean wingtip velocity $\bar{u}_{tip} = 1~[LT^{-1}]$ by definition from eqn.~\ref{eqn:revolving_angle}, the fluid viscosity $\nu = 1.477 \cdot 10^{-4}~[L^2 T^{-1}]$ and the mean chord, the ratio between the wing surface area $A$ and the wing length $R_w$, $c_m = A/R_w = 0.304~[L]$. This yields $Re = 2060$. Additionally, the lift and drag coefficient are defined as below 
\begin{equation}
    C_L = \frac{F_L}{M L T^{-2}} \text{; } C_D = \frac{F_D}{M L T^{-2}}
\end{equation}
where the lift $F_L$ is the force in the vertical direction $Oz$ and the drag $F_D$ is the force perpendicular to the plane formed by the vertical and the wing spanwise axes, as shown in figure~\ref{fig:revolving_wing_config}a.

\begin{figure}[th]
\centering
\includegraphics[width=0.5\textwidth]{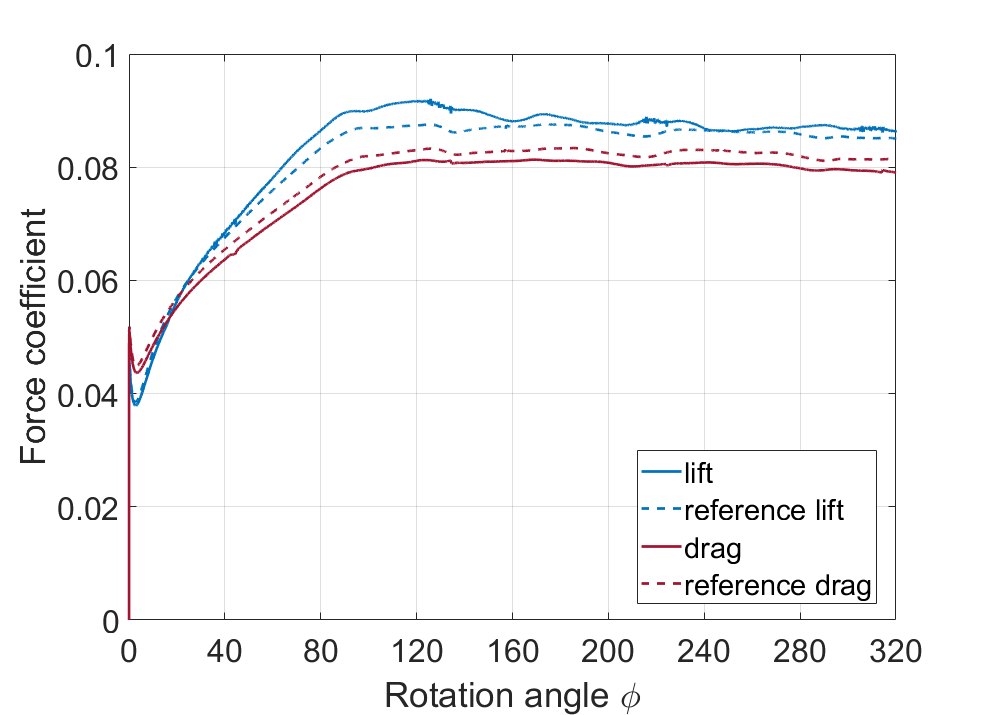}
\caption{Comparison of lift and drag coefficients for a rigid wing, calculated using the coupling between FLUSI and the developed mass-spring solver, with the reference data from~\cite{RigidRevolWing}.}
\label{fig:rigid_wing_lift_drag}
\end{figure}

The computed lift and drag coefficients are shown in figure~\ref{fig:rigid_wing_lift_drag} along with the reference values from~\cite{RigidRevolWing}. To evaluate quantitatively the error, the average lift and drag during the steady state (for rotation angles $\phi$ varying from $160^\circ$ to $320^\circ$) are computed and compared with the reference. A very good agreement is obtained with the relative error of $1.3 \%$ for the drag and $1.6 \%$ for the lift. 

From the results obtained from these two test cases, the satisfactory agreements can give us the confidence about the solid solver, based on mass-spring system, as well as the coupling with the flow solver FLUSI. Any difference between all the numerical studies carried out can be explained by the difference between the continuum model and the discrete model together with the way of generating the mask function.
\section{Results and discussion}\label{sec:results}


In the following we present results of high resolution computations of revolving bumblebee wings which are either rigid, flexible or highly flexible. 
First we perform computations for different resolutions to check the mesh convergence for both fluid and solid solvers. Then a comparison of the flexible wings with the rigid case allows to assess the influence of the wing deformation on the aerodynamic forces. The setup is exactly the same as described in the revolving wing test case of the previous section. The only difference is the wing shape which is now changed back to the one presented in section~\ref{sec:Wing_structure}. As a result, the length scale and the mass scale are changed as follows: $L = 15~mm$ and $M = \rho_{air} \times L^3 = 4.13~mg$ while the time scale remains the same $T = 1~s$. The corresponding Reynolds number is $1800$ where the fluid viscosity is assumed to be $\nu = 1.477 \cdot 10^{-4}~[L^2 T^{-1}]$, the wing tip velocity $u_{tip} = 1~[L T^{-1}]$ and the mean chord calculated from the new wing surface area is $c_m = A/R_w = 0.266~[L]$. 

\subsection{Study of mesh convergence}

\subsubsection{Fluid mesh}

The following mesh convergence study for the fluid solver is performed considering five different resolutions: $128 \times 128 \times 64$, $256 \times 256 \times 128$, $512 \times 512 \times 256$, $768 \times 768 \times 384$ and $1024 \times 1024 \times 512$. 

\begin{figure}[ht]
\centering
    \includegraphics[width=0.5\textwidth]{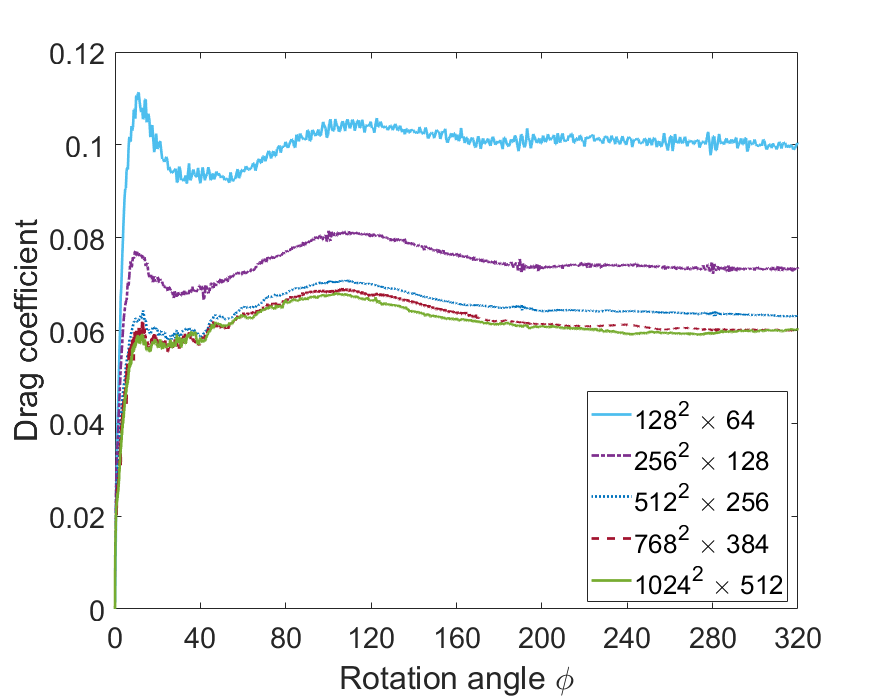}\\

  \caption{Drag coefficient generated by a flexible wing, calculated at different resolutions: $128^2 \times 64$, $256^2 \times 128$, $512^2 \times 256$, $768^2 \times 384$ and $1024^2 \times 512$ .}
  \label{fig:convergence_study_BB_revolving}
\end{figure}

\begin{figure}[th]
\centering
\includegraphics[width=0.40\textwidth]{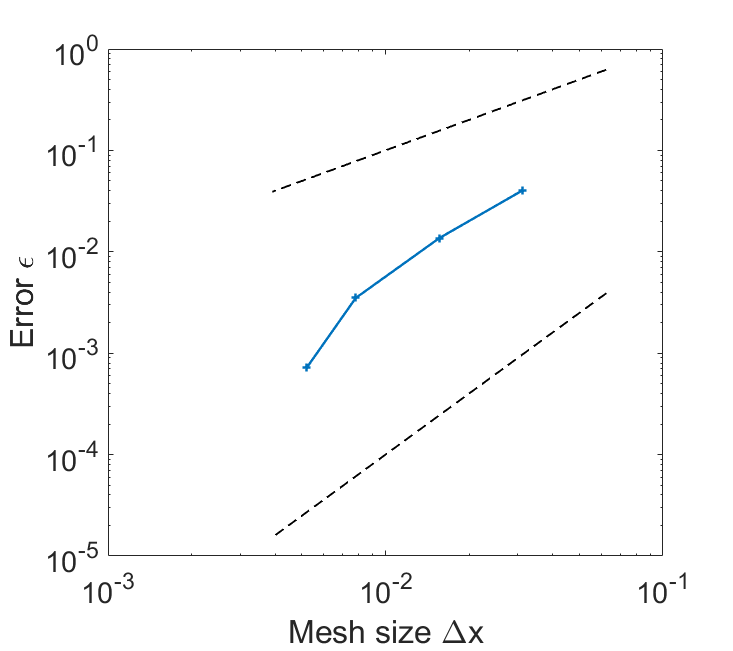}
\caption{The error of the mean drag versus mesh size. The dashed lines represent first and second order convergence.}
\label{fig:convergence_study_BB_revolving_error}
\end{figure}

The mean drag generated during the second half cycle of the rotation is chosen for the evaluation of the mesh convergence ($160^\circ \le \phi \le 320^\circ$). Because it is impossible to obtain the exact values for the mean drag in this case, we use here the result obtained with the finest mesh as a reference value. The relative error of the mean drag with respect to the reference drag for all the mesh size is shown in figure~\ref{fig:convergence_study_BB_revolving_error}. In all the simulations, the penalization parameter $C_{\eta}$ is chosen to satisfy that the number of points per thickness of the penalization boundary layer $K_{\eta} =\sqrt{\nu C_{\eta}}/ \Delta x$ is always constant (as recommended in~\cite{ThomasThesis}) and equal to $0.052$. The drag obtained for each simulation (figure~\ref{fig:convergence_study_BB_revolving}) shows the convergence to the finest resolution solution when we refine the mesh. The spatial convergence exhibits a first to second order behavior when we plot the error versus the mesh size.

\subsubsection{Solid mesh}

\begin{figure}[th]
\centering
\includegraphics[width=0.50\textwidth]{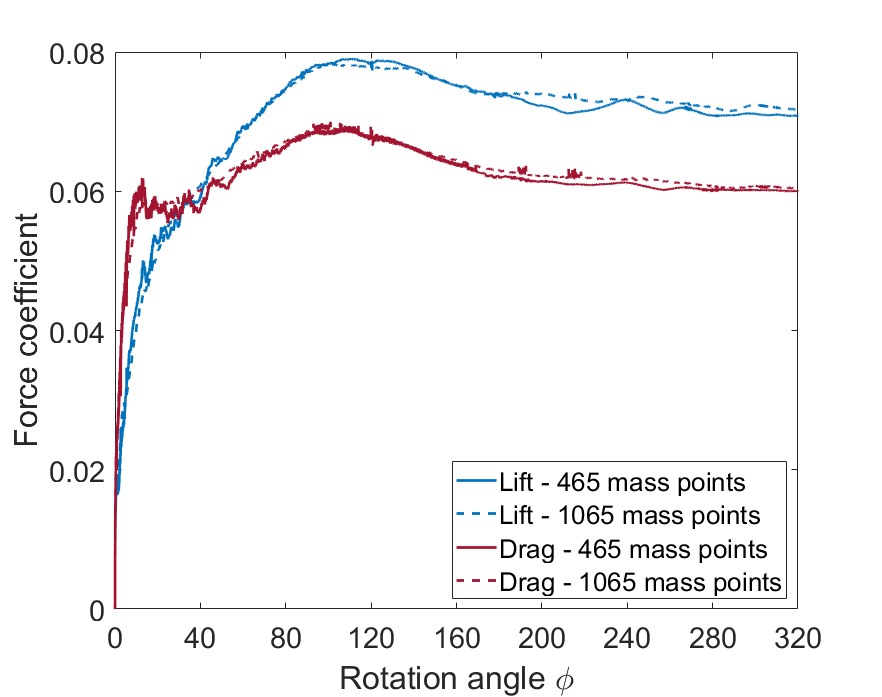}
\caption{Lift and drag coefficients generated by a revolving flexible wing discretized by 465 and 1065 mass points at $Re= 1800$.}
\label{fig:convergence_study_BB_revolving_solid}
\end{figure}

As mentioned above, the dynamics of the mass-spring system depends strongly on the mesh size. Thus another convergence test on the number of mass points is performed. Two simulations of a revolving flexible wing at resolution $768^2 \times 384$ are run to compare between a medium-mesh and a fine-mesh wing which are discretized by $465$ and $1065$ mass points, respectively. As shown here in figure \ref{fig:convergence_study_BB_revolving_solid}, although the number of mass points is more than doubled, the forces remain almost unchanged with an increase of $1.1 \%$ and $0.8 \%$ in average lift and drag coefficients during the steady state, respectively. Since the fluid solver is itself already costly in term of CPU time, the medium-mesh wing with $465$ mass points is sufficient and can be chosen for the following study in section \ref{subsec:Flexibility_Influence}.

\subsection{Influence of wing flexibility}
\label{subsec:Flexibility_Influence}

To examine the influence of vein stiffness on the aerodynamic performance of the wing, the flexural rigidity of veins will be varied by changing the Young modulus $E$. Two values of the Young modulus are used: $E= 1.25 \cdot 10^{8}~[ML^{-1}T^{-2}]$ and $E=1.25 \cdot 10^{7}~[ML^{-1}T^{-2}]$, corresponding to the flexible and highly flexible cases, respectively. 

\begin{figure*}[ht]
\centering
  \begin{tabular}{@{}c@{}}
     \includegraphics[width=1\textwidth]{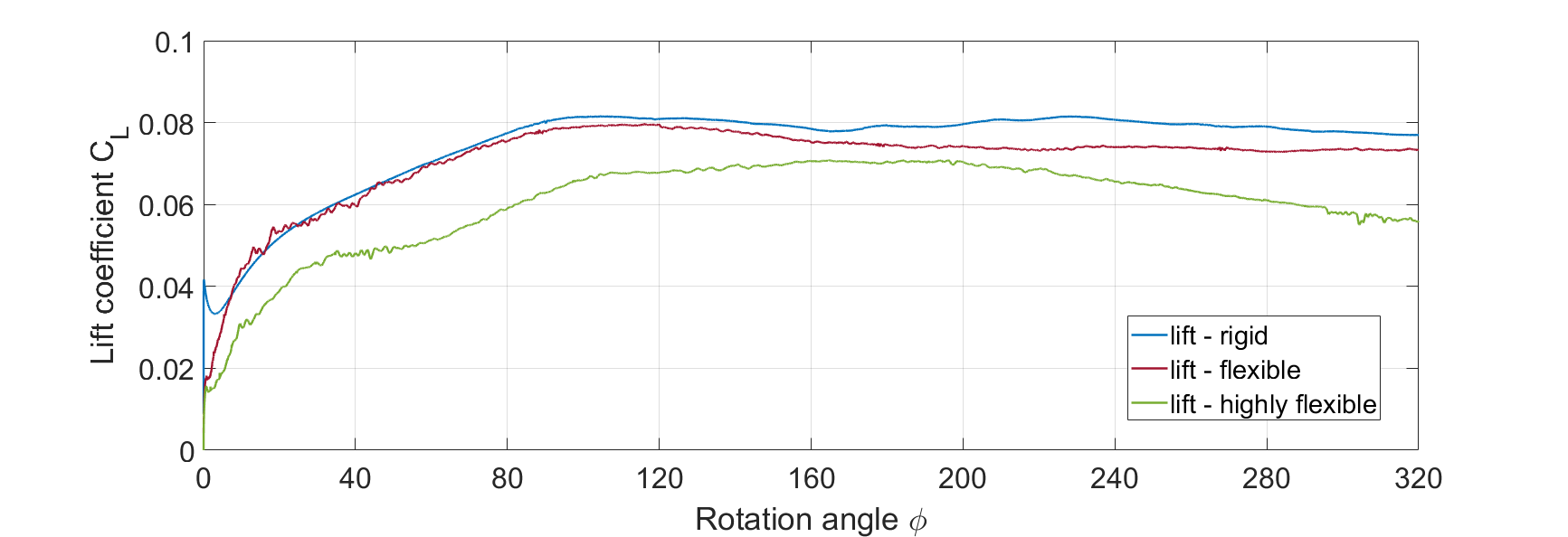}\\
    \small (a) Lift coefficients.
  \end{tabular}

  \begin{tabular}{@{}c@{}}
    \includegraphics[width=1\textwidth]{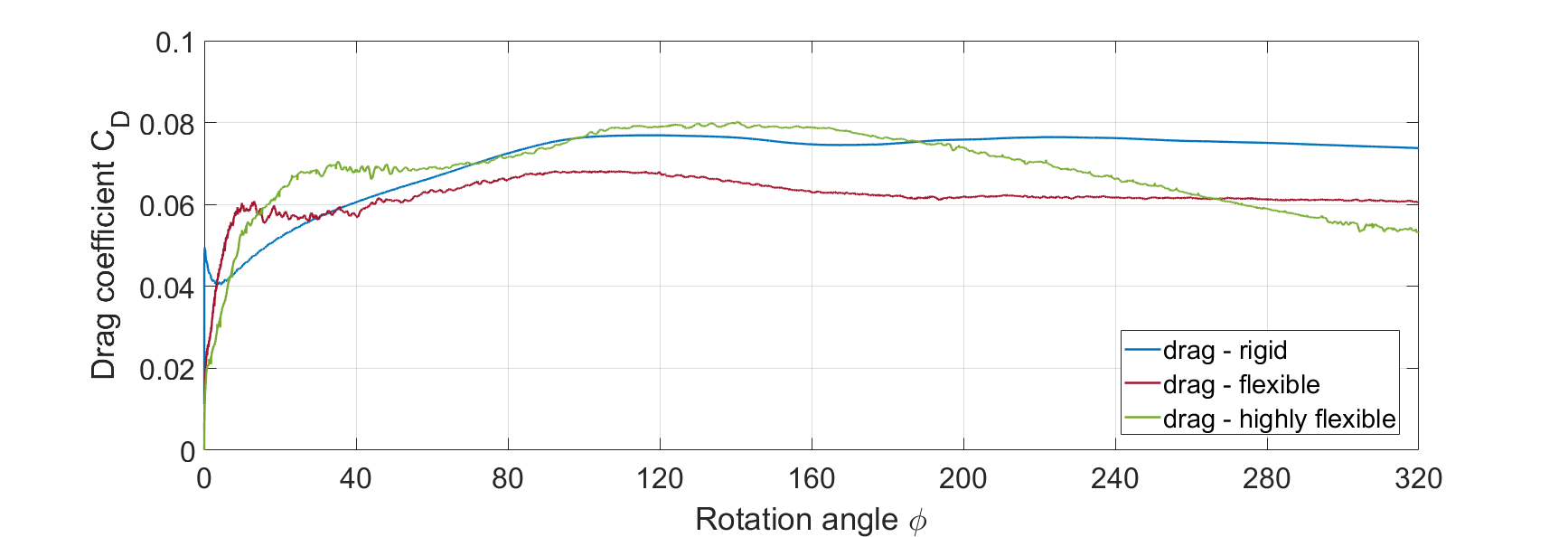}\\
    \small (b) Drag coefficient.
  \end{tabular}
  \caption{Influence of wing flexibility on lift and drag coefficients of a revolving wing at $Re= 1800$.}
  \label{fig:flexibility_influence}
\end{figure*}

Lift and drag coefficients at resolution $1024^2 \times 512$ for the rigid, flexible and highly flexible cases are presented in figure~\ref{fig:flexibility_influence}.

During the transition phase (rotation angle $\phi \le 40^\circ$), the lift generated by the rigid wing increases instantly and then decreases before going up again. The drag follows the same trend as the lift, but is larger in magnitude. When the flexibility of the wing is taken into account, the rapid rise at the beginning of the forces for both flexible and highly flexible wings disappear. Instead, the forces increase gradually and the more flexible the wing is, the lower the lift and the higher the drag are.

At steady state, similar behaviors between the rigid and the flexible wings can be observed. When the rotation angle reaches $160^\circ$, the forces generated by these two wings are stabilized. This can be explained by the fact that no dynamic deformation of the wings takes place and just the shape plays a role.

\begin{figure}[ht]
\centering
\includegraphics[scale=0.40]{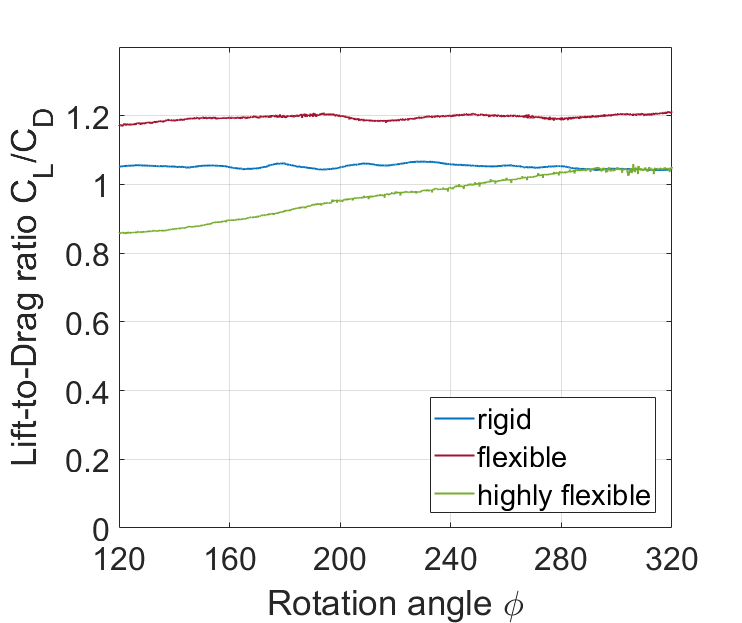}
\caption{Lift-to-drag ratio for the three wings during the steady state, rotation angle $120^\circ \le \phi \le 320^\circ$.}
\label{fig:lift_to_drag}
\end{figure}

We also find that the lift-to-drag ratio at the steady state of the flexible wing is $1.2$, $14.5 \%$ higher than the one of the rigid case, which is only $1.05$ (figure~\ref{fig:lift_to_drag}). This finding is consistent with conclusions found in literature \cite{FlexInfluenceZhao,FlexInfluenceMeerendonk}. A flexible wing generates less lift and drag than a rigid one. However, due to the flexibility of the wing, the bending in the chordwise direction makes the effective geometric angle of attack decrease and alters the direction of the total resultant force upward \cite{FlexInfluenceMeerendonk}. This makes the lift-to-drag ratio raise and allows better flight performance.

\begin{figure}[ht]
\centering
  \begin{tabular}{@{}c@{}}
     \includegraphics[width=0.45\textwidth]{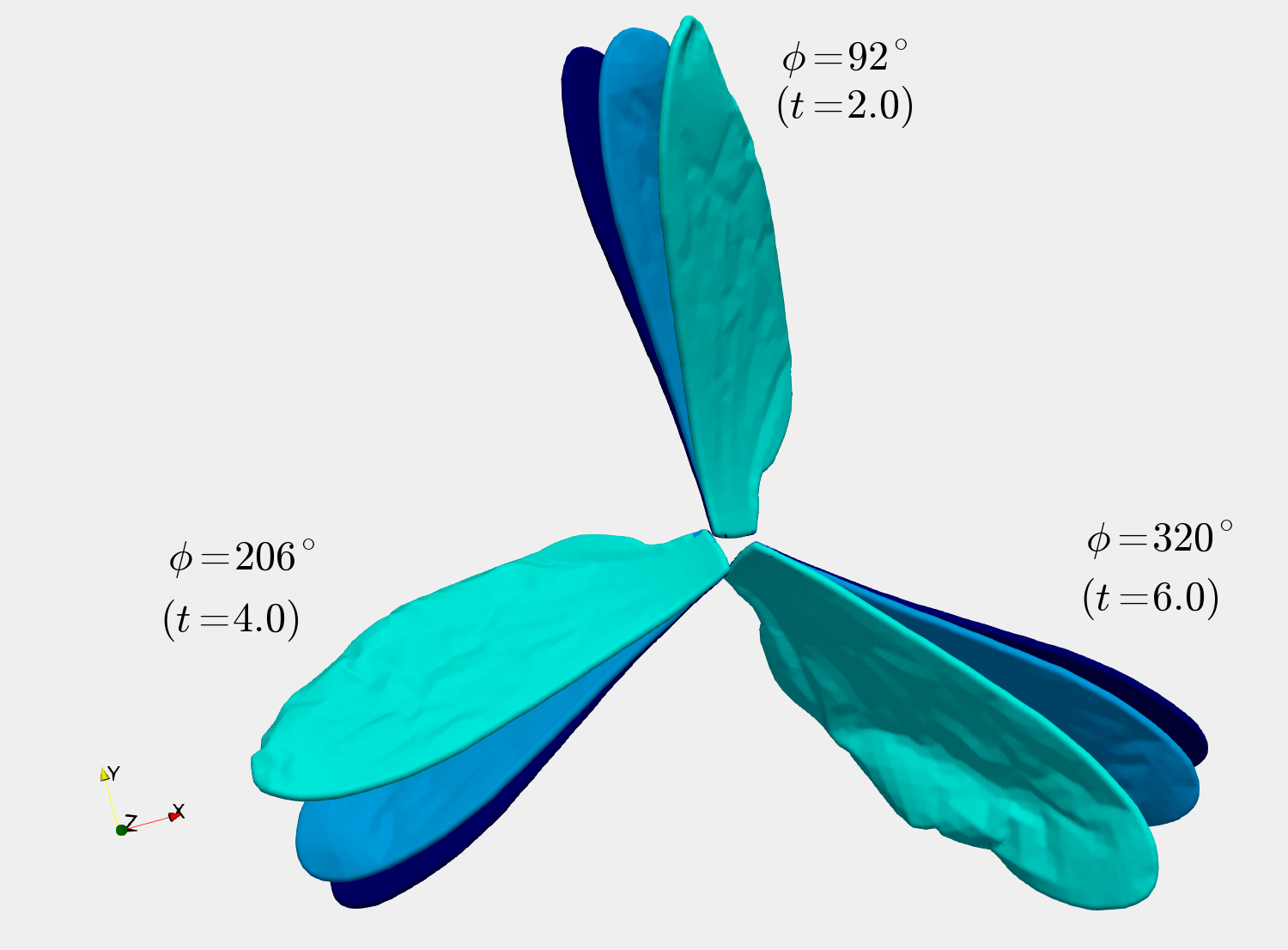}\\
    \small (a) Top view.
  \end{tabular}

  \begin{tabular}{@{}c@{}}
    \includegraphics[width=0.45\textwidth]{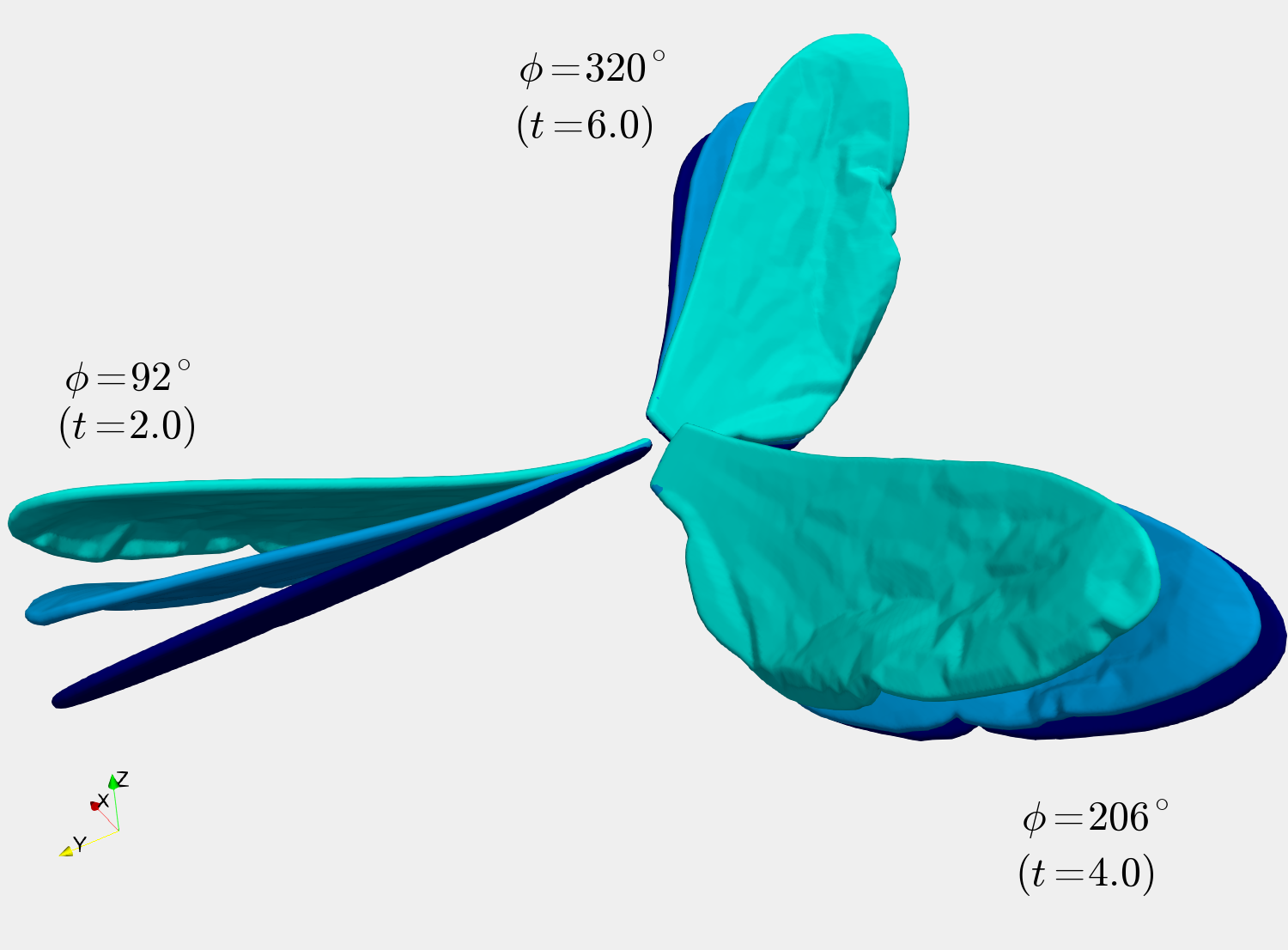}\\
    \small (b) Side view.
  \end{tabular}
  \caption{Wing deformation corresponding to rigid (dark blue), flexible (blue) and highly flexible (light blue) wings at three time instants, $t=2$, $t=4$ and $t=6$.}
  \label{fig:mask_deformation_all_time}
\end{figure}

On the contrary, the highly flexible wing acts differently. Both the lift and the drag increase gradually to attain their maximum values at the rotation angle $\phi = 120^\circ$ and then decline instead of being stabilized as in the other simulations. The lift-to-drag ratio is surprisingly much less than the one of the rigid case at the beginning of the steady state but then increases and keeps up with the rigid wing. This can be explained by the fact that the bending of the wing in the spanwise direction (figure~\ref{fig:mask_deformation_all_time}) prevents the development of the LEV growing further toward the wing tip and makes the LEV burst sooner at mid-span of the wing.

The change of aerodynamic forces compared to the rigid case is linked to the deformation of the wing, which is modeled by the mass-spring solver. The wing deformation for all three cases is shown in the same figure~\ref{fig:mask_deformation_all_time} for comparison at three time instants $t=2$, $t=4$ and $t=6$. By applying the functional approach, the difference between the vein and the membrane is visible in the visualization. 

\begin{figure*}[ht]
\begin{subfigure}{.5\textwidth}
  \centering
  \includegraphics[width=0.9\linewidth]{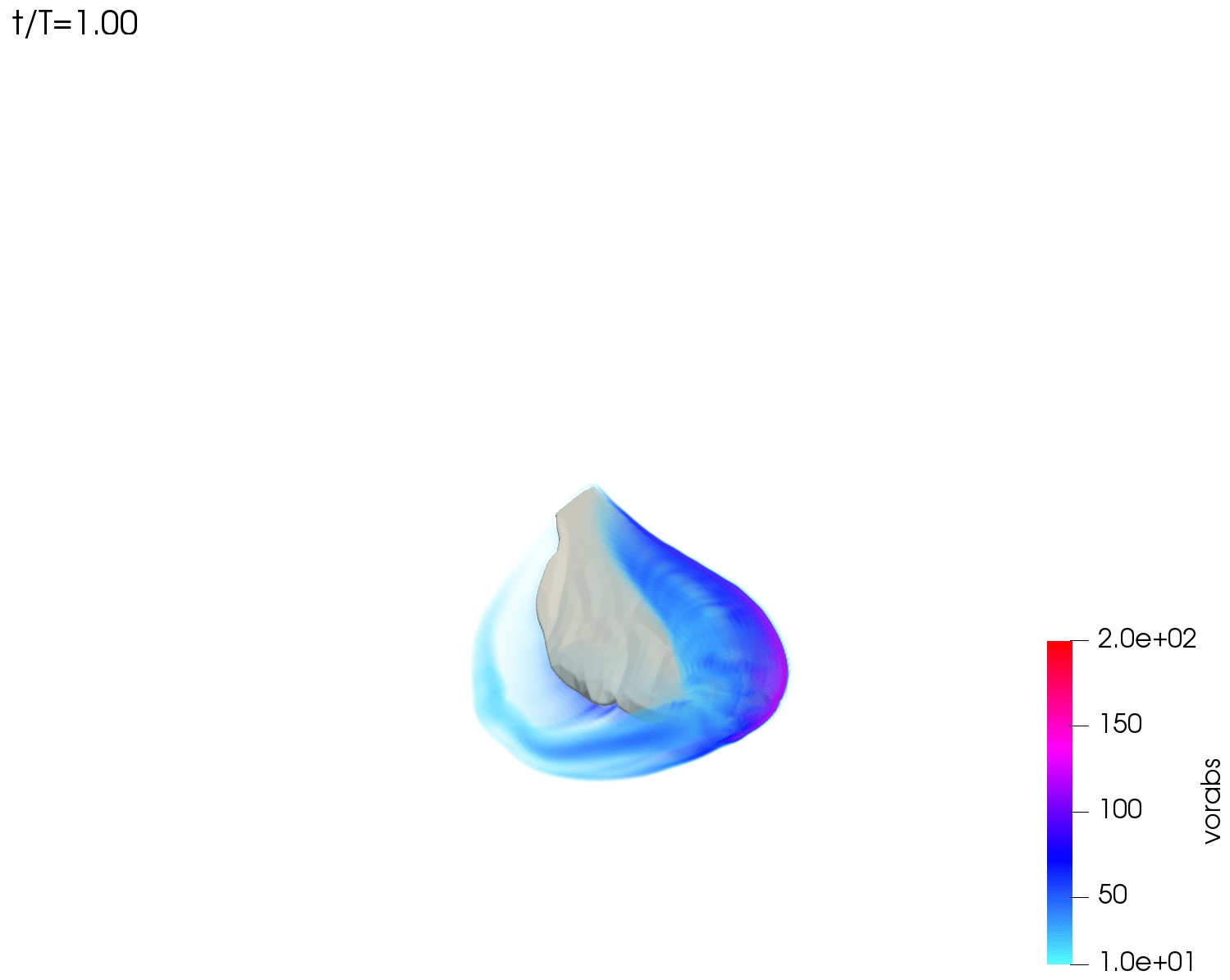}  
  \label{fig:sub-first}
\end{subfigure}
\begin{subfigure}{.5\textwidth}
  \centering
  \includegraphics[width=0.9\linewidth]{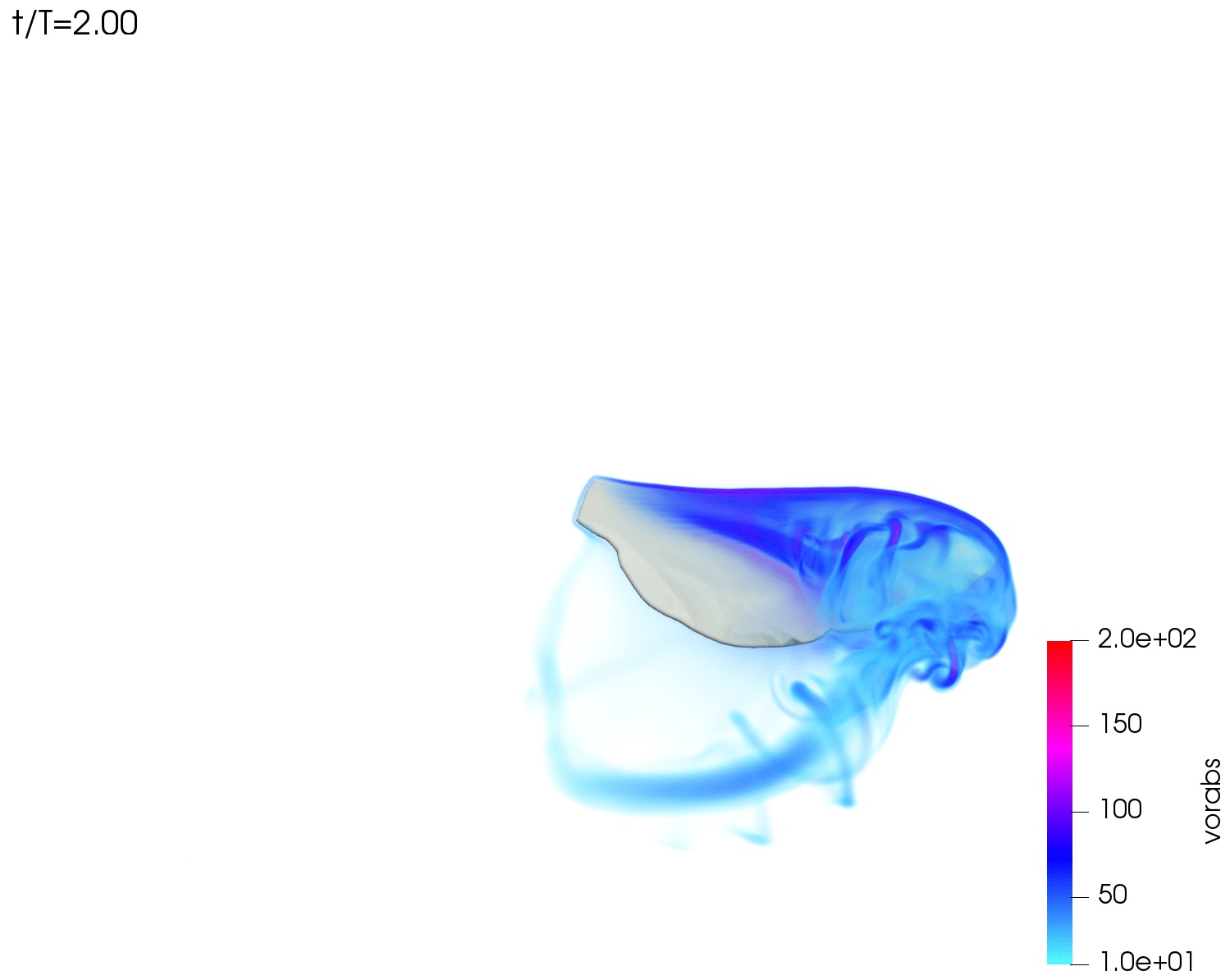} 
  \label{fig:sub-second}
\end{subfigure}

\begin{subfigure}{.5\textwidth}
  \centering
  \includegraphics[width=0.9\linewidth]{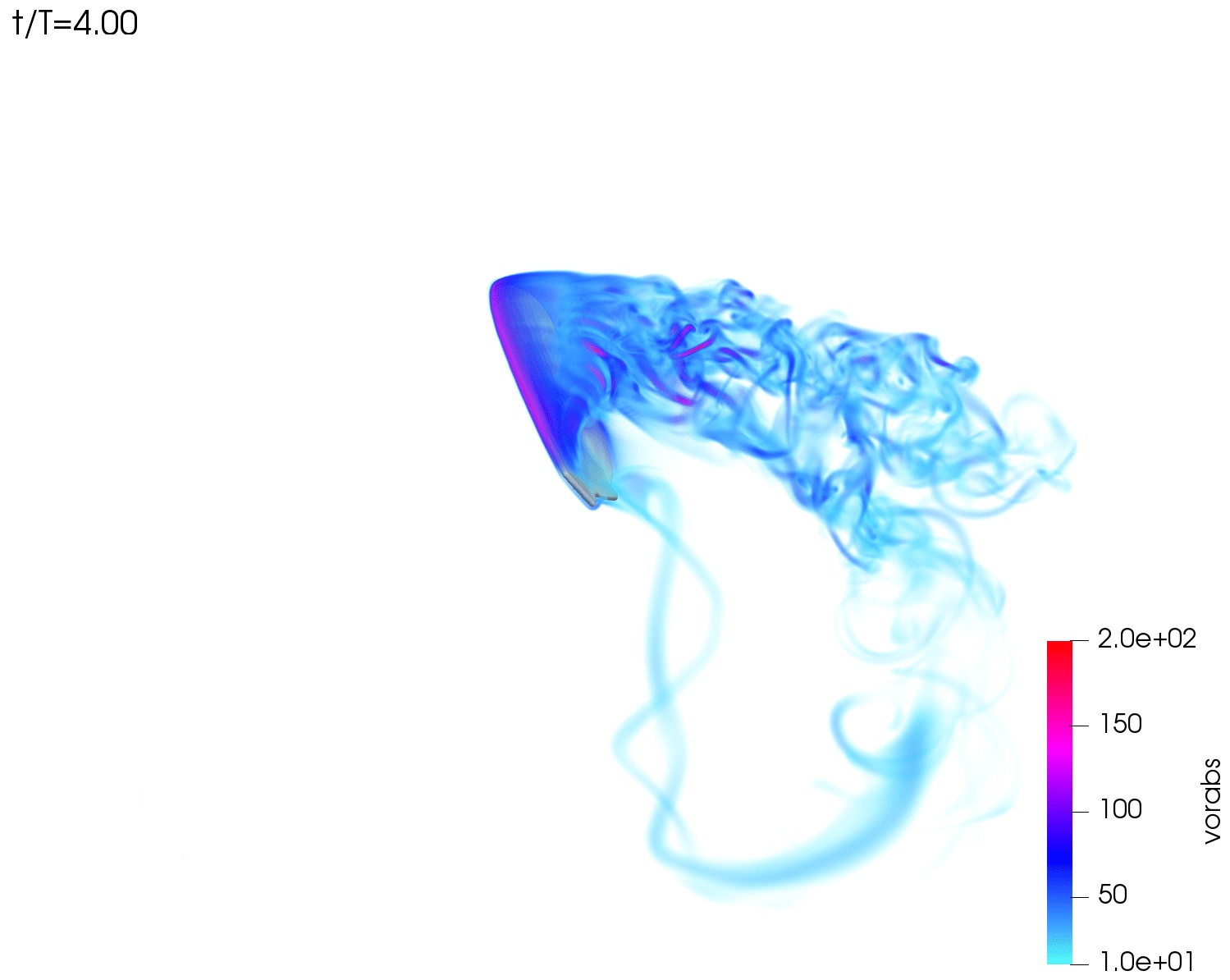} 
  \label{fig:sub-third}
\end{subfigure}
\begin{subfigure}{.5\textwidth}
  \centering
  \includegraphics[width=0.9\linewidth]{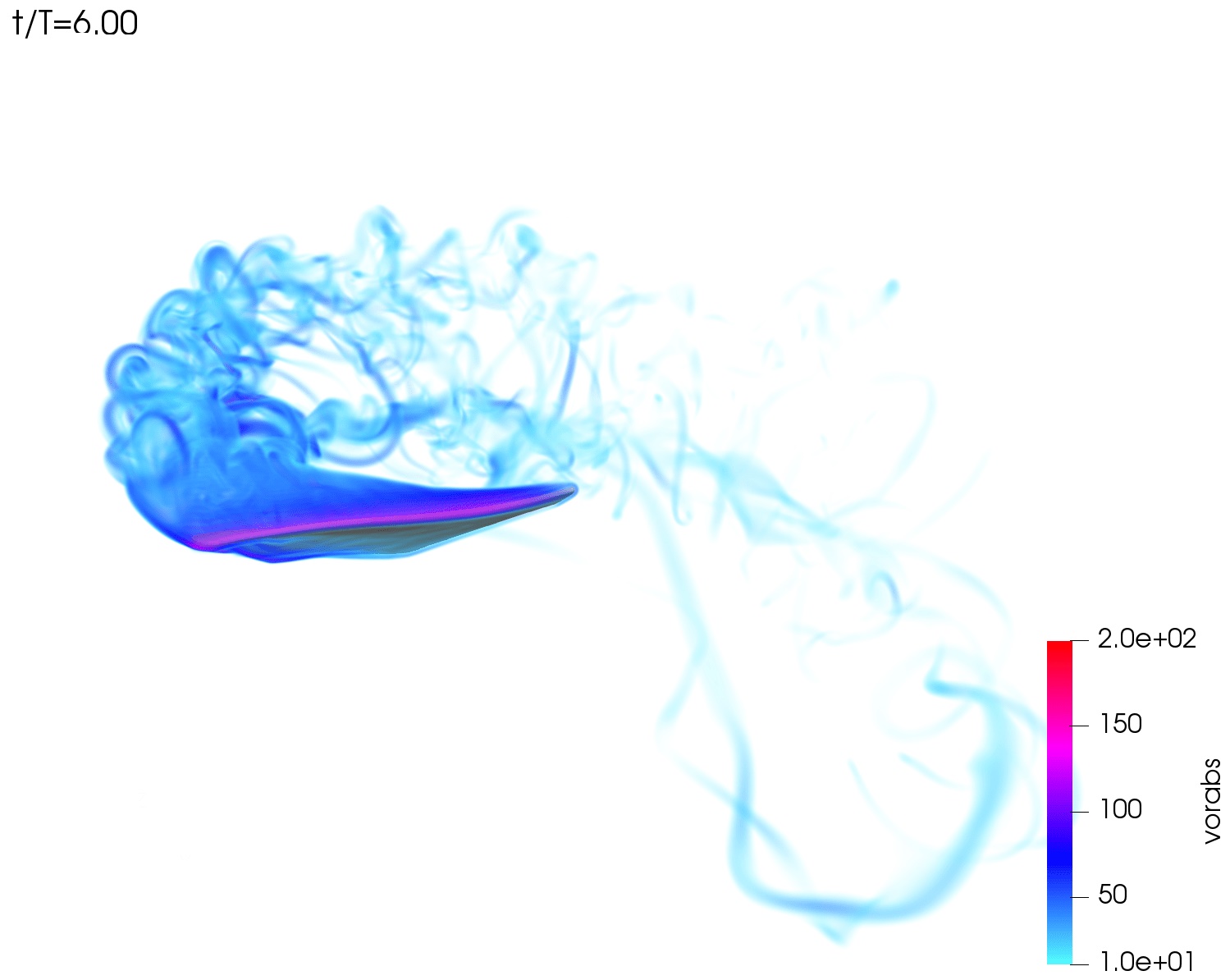}
  \label{fig:sub-fourth}
\end{subfigure}
\caption{Flows generated by a flexible revolving wing, visualized by their vorticity magnitude $|\mathbf{\omega}|$ at four
time instants $t=1,2,4$ and $6$. The simulation is performed with resolution $1024^2 \times 512$.}
\label{fig:BB-rev_flexible_vorabs_10-200}
\end{figure*}

At the finest resolution of the mesh ($1024^2 \times 512$), the flows generated by the flexible wing are shown (figure~\ref{fig:BB-rev_flexible_vorabs_10-200}) by plotting their vorticity magnitude at four time instants of the simulation. The formations of the leading edge vortex as well as the tip vortex can be observed clearly at the beginning of the rotation ($t=1.0$ and $t=2.0$). Then, the vortex burst happens and a region of inhomogeneous vorticity forms at the wing tip. However, the LEV remains attached to the wing surface and this results in constant lift and drag.
 
\section{Conclusions}\label{sec:conclusions}

%
We presented a numerical approach for fluid-structure interaction in the open source framework FLUSI, which is based on a mass-spring model describing the structure of the insect wings and a pseudospectral method for solving the incompressible Navier--Stokes equations. For imposing no-slip boundary conditions in the complex time-changing geometry we used the volume penalization technique. The solver has been implemented on massively parallel supercomputers using MPI and allows high resolution computations, here with more than half a billion grid points. Code validation for two classical benchmarks, a flow past a cylinder with a flexible appendage and the flow generated by a rigid revolving wing, is likewise presented.

Considering the flexible wing, the flexibility reduces the buildup of the aerodynamic force during the beginning of motion. Nevertheless, after the start-up phase, the wing yields a steady state configuration, and no significant oscillation nor unsteady deformation of the wing are observed. A better aerodynamic performance of the flexible wing, characterized by the increase of the lift-to-drag ratio during the steady state, is explained by the decrease of the effective angle of attack caused by the deformation of the flexible wing. On the other hand, the highly flexible wing appears to be less efficient than the rigid wing. This can be interpreted that there is an optimized zone of wing flexibility, which is ideal for flying.

For flapping wings we anticipate that flexibility will become important because of the dynamic wing deformation. In the near future we are planing high resolution numerical simulations of flapping insect flight with flexible wings where the dynamical deformation plays an important role. 

The limitations of the current approach are the resolution and CPU time requirements imposed by the use of an uniform grid. Hence large scale simulations become prohibitively expensive. Moreover, the thickness of bumblebee wing is much smaller than the spatial mesh size in our present simulations. Consequently, the virtual thickness of wings studied here is set to $4$ times the mesh size, necessary for the usage of the volume penalization method.

An adaptive version of the FLUSI code, likewise fully parallel, using wavelet-based grid refinement is currently being developed to be able to reduce memory and CPU time requirements. High resolution numerical simulation of flapping flight for larger species and large Reynolds numbers will thus become possible. 

Implementing the solid solver presented in the current paper into the adaptive Navier--Stokes solver will allow to perform fluid-structure interaction on adaptive grids at reduced computational cost. 

\section*{Acknowledgments}
Financial support from the Agence Nationale de la Recher-che (ANR Grant No. 15-CE40-0019)
and Deutsche Forschungsgemeinschaft (DFG Grant No. SE 824/26-1), project AIFIT, is gratefully acknowledged.
The authors were granted access to the HPC resources of IDRIS under the Allocation No. 2018-91664 attributed by GENCI (Grand \'Equipement National de Calcul Intensif). For this work, we were also granted access to the HPC resources of Aix-Marseille Universit\'e financed by the project Equip{@}Meso (No. ANR-10-EQPX- 29-01).
The authors thankfully acknowledge financial support granted by the minist\`eres des Affaires \'etrang\`eres et du d\'eveloppement international (MAEDI) et de l'Education nationale et l'enseignement sup\'erieur, de la recherche et de l'innovation (MENESRI), and the Deutscher Akademischer Austauschdienst (DAAD) within the French-German Procope project FIFIT.
D.K. gratefully acknowledges financial support from the JSPS KAKENHI Grant No. JP18K13693.
K.S. thanks the organizers of ICCFD10, in particular C.H. Bruneau, for inviting him for a plenary talk.


\bibliographystyle{model1-num-names}
\bibliography{references}

\begin{thebibliography}{50}
\expandafter\ifx\csname natexlab\endcsname\relax\def\natexlab#1{#1}\fi
\providecommand{\url}[1]{\texttt{#1}}
\providecommand{\href}[2]{#2}
\providecommand{\path}[1]{#1}
\providecommand{\DOIprefix}{doi:}
\providecommand{\ArXivprefix}{arXiv:}
\providecommand{\URLprefix}{URL: }
\providecommand{\Pubmedprefix}{pmid:}
\providecommand{\doi}[1]{\href{http://dx.doi.org/#1}{\path{#1}}}
\providecommand{\Pubmed}[1]{\href{pmid:#1}{\path{#1}}}
\providecommand{\bibinfo}[2]{#2}
\ifx\xfnm\relax \def\xfnm[#1]{\unskip,\space#1}\fi
\bibitem[{Young et~al.(2009)Young, Walker, Bomphrey, Taylor, and
  Thomas}]{AeroYoung}
\bibinfo{author}{J.~Young}, \bibinfo{author}{S.~M. Walker},
  \bibinfo{author}{R.~J. Bomphrey}, \bibinfo{author}{G.~K. Taylor},
  \bibinfo{author}{A.~L.~R. Thomas},
\newblock \bibinfo{title}{Details of insect wing design and deformation enhance
  aerodynamic function and flight efficiency},
\newblock \bibinfo{journal}{Science} \bibinfo{volume}{325}
  (\bibinfo{year}{2009}) \bibinfo{pages}{1549--1552}.
\bibitem[{Nakata and Liu(2012)}]{FSINakata2012}
\bibinfo{author}{T.~Nakata}, \bibinfo{author}{H.~Liu},
\newblock \bibinfo{title}{A fluid-structure interaction model of insect flight
  with flexible wings},
\newblock \bibinfo{journal}{Journal of Computational Physics}
  \bibinfo{volume}{231} (\bibinfo{year}{2012}) \bibinfo{pages}{1822--1847}.
\bibitem[{Combes and Daniel(2003)}]{CombesI}
\bibinfo{author}{S.~A. Combes}, \bibinfo{author}{T.~L. Daniel},
\newblock \bibinfo{title}{Flexural stiffness in insect wings {I}. {S}caling and
  the influence of wing venation},
\newblock \bibinfo{journal}{The Journal of Experimental Biology}
  \bibinfo{volume}{206} (\bibinfo{year}{2003}) \bibinfo{pages}{2979--2987}.
\bibitem[{Engels et~al.(2016)Engels, Kolomenskiy, Schneider, and
  Sesterhenn}]{Flusi}
\bibinfo{author}{T.~Engels}, \bibinfo{author}{D.~Kolomenskiy},
  \bibinfo{author}{K.~Schneider}, \bibinfo{author}{J.~Sesterhenn},
\newblock \bibinfo{title}{Flusi: A novel parallel simulation tool for flapping
  insect flight using a {F}ourier method with volume penalization},
\newblock \bibinfo{journal}{SIAM J. Sci. Comp.} \bibinfo{volume}{38}
  (\bibinfo{year}{2016}) \bibinfo{pages}{S03--S24}.
\bibitem[{Kolomenskiy et~al.(2014)Kolomenskiy, Elimelech, and
  Schneider}]{KES14}
\bibinfo{author}{D.~Kolomenskiy}, \bibinfo{author}{Y.~Elimelech},
  \bibinfo{author}{K.~Schneider},
\newblock \bibinfo{title}{Leading-edge vortex shedding from rotating wings},
\newblock \bibinfo{journal}{Fluid Dyn. Res.} \bibinfo{volume}{46}
  (\bibinfo{year}{2014}) \bibinfo{pages}{031421}.
\bibitem[{Engels et~al.(2018)Engels, Kolomenskiy, Schneider, Farge, Lehmann,
  and Sesterhenn}]{RigidRevolWing}
\bibinfo{author}{T.~Engels}, \bibinfo{author}{D.~Kolomenskiy},
  \bibinfo{author}{K.~Schneider}, \bibinfo{author}{M.~Farge},
  \bibinfo{author}{F.-O. Lehmann}, \bibinfo{author}{J.~Sesterhenn},
\newblock \bibinfo{title}{Helical vortices generated by flapping wings of
  bumblebees},
\newblock \bibinfo{journal}{Fluid Dyn. Res.} \bibinfo{volume}{50}
  (\bibinfo{year}{2018}) \bibinfo{pages}{011419}.
\bibitem[{Engels et~al.(2019)Engels, Kolomenskiy, Schneider, Farge, Lehmann,
  and Sesterhenn}]{EKSFLS19}
\bibinfo{author}{T.~Engels}, \bibinfo{author}{D.~Kolomenskiy},
  \bibinfo{author}{K.~Schneider}, \bibinfo{author}{M.~Farge},
  \bibinfo{author}{F.~Lehmann}, \bibinfo{author}{J.~Sesterhenn},
\newblock \bibinfo{title}{The impact of turbulence on flying insects in
  tethered and free flight: high-resolution numerical experiments},
\newblock \bibinfo{journal}{Phys. Rev. Fluids} \bibinfo{volume}{4}
  (\bibinfo{year}{2019}) \bibinfo{pages}{013103}.
\bibitem[{Jarrousse(2014)}]{Jar12}
\bibinfo{author}{O.~Jarrousse}, \bibinfo{title}{Modified mass-spring system for
  physically based deformation modeling}, \bibinfo{publisher}{KIT Scientific
  Publishing}, \bibinfo{year}{2014}.
\bibitem[{Terzopoulos et~al.(1987)Terzopoulos, Platt, Barr, and
  Fleischer}]{MembraneTerzopoulos}
\bibinfo{author}{D.~Terzopoulos}, \bibinfo{author}{J.~Platt},
  \bibinfo{author}{A.~H. Barr}, \bibinfo{author}{K.~Fleischer},
\newblock \bibinfo{title}{Elastically deformable models},
\newblock \bibinfo{journal}{ACM Siggraph Computer Graphics}
  \bibinfo{volume}{21} (\bibinfo{year}{1987}) \bibinfo{pages}{205--214}.
\bibitem[{Eischen et~al.(1996)Eischen, Deng, and Clapp}]{MembraneEischen}
\bibinfo{author}{J.~W. Eischen}, \bibinfo{author}{S.~Deng},
  \bibinfo{author}{T.~G. Clapp},
\newblock \bibinfo{title}{Finite element modeling and control of flexible
  fabric parts},
\newblock \bibinfo{journal}{IEEE Computer Graphics and Applications}
  \bibinfo{volume}{16} (\bibinfo{year}{1996}) \bibinfo{pages}{71--80}.
\bibitem[{Cai et~al.(2016)Cai, Lin, and Lee}]{Cai16}
\bibinfo{author}{J.~Cai}, \bibinfo{author}{F.~Lin}, \bibinfo{author}{Y.~Lee},
  \bibinfo{title}{Modeling and dynamics simulation for deformable objects of
  orthotropic materials}, \bibinfo{publisher}{Springer Berlin Heidelberg},
  \bibinfo{year}{2016}.
\bibitem[{Nealen et~al.(2006)Nealen, M\"uller, Keiser, Boxerman, and
  Carlson}]{DeforModelNealen}
\bibinfo{author}{A.~Nealen}, \bibinfo{author}{M.~M\"uller},
  \bibinfo{author}{R.~Keiser}, \bibinfo{author}{E.~Boxerman},
  \bibinfo{author}{M.~Carlson},
\newblock \bibinfo{title}{Physically based deformable models in computer
  graphics},
\newblock \bibinfo{journal}{Computer Graphics forum} \bibinfo{volume}{25}
  (\bibinfo{year}{2006}) \bibinfo{pages}{809--836}.
\bibitem[{Miller and Peskin(2009)}]{PeskinClapandFling}
\bibinfo{author}{L.~A. Miller}, \bibinfo{author}{C.~S. Peskin},
\newblock \bibinfo{title}{Flexible clap and fling in tiny insect flight},
\newblock \bibinfo{journal}{J. Exp. Biol} \bibinfo{volume}{212}
  (\bibinfo{year}{2009}) \bibinfo{pages}{3076–3090}.
\bibitem[{Yeh and Alexeev(2016)}]{FSIYeh2016}
\bibinfo{author}{P.~Yeh}, \bibinfo{author}{A.~Alexeev},
\newblock \bibinfo{title}{Biomimetic flexible plate actuators are faster and
  more efficient with a passive attachment},
\newblock \bibinfo{journal}{Acta Mechanica Sinica} \bibinfo{volume}{32}
  (\bibinfo{year}{2016}) \bibinfo{pages}{1001--1011}. \bibinfo{note}{Doi:
  \url{10.1007/s10409-016-0592-0}}.
\bibitem[{Lentink and Dickinson(2009)}]{RevolWingLentink09}
\bibinfo{author}{D.~Lentink}, \bibinfo{author}{M.~H. Dickinson},
\newblock \bibinfo{title}{Rotational accelerations stabilize leading edge
  vortices on revolving fly wings},
\newblock \bibinfo{journal}{Journal of Experimental Biology}
  \bibinfo{volume}{212} (\bibinfo{year}{2009}) \bibinfo{pages}{2705--2719}.
  \bibinfo{note}{Doi: \url{10.1242/jeb.022269}}.
\bibitem[{Jardin and David(2014)}]{RevolvWingJardin14}
\bibinfo{author}{T.~Jardin}, \bibinfo{author}{L.~David},
\newblock \bibinfo{title}{Spanwise gradients in flow speed help stabilize
  leading-edge vortices on revolving wings},
\newblock \bibinfo{journal}{Phys. Rev. E} \bibinfo{volume}{90}
  (\bibinfo{year}{2014}) \bibinfo{pages}{013011}. \bibinfo{note}{Doi:
  \url{10.1103/PhysRevE.90.013011}}.
\bibitem[{Jones et~al.(2016)Jones, Medina, Spooner et~al.}]{RevolWingJones16}
\bibinfo{author}{A.~Jones}, \bibinfo{author}{A.~Medina},
  \bibinfo{author}{H.~Spooner}, et~al.,
\newblock \bibinfo{title}{Biomimetic flexible plate actuators are faster and
  more efficient with a passive attachment},
\newblock \bibinfo{journal}{Exp Fluids} \bibinfo{volume}{57}
  (\bibinfo{year}{2016}) \bibinfo{pages}{1--16}. \bibinfo{note}{Doi:
  \url{10.1007/s00348-016-2143-7}}.
\bibitem[{Di et~al.(2017)Di, Kolomenskiy, Nakata, and Liu}]{RevolWingDi18}
\bibinfo{author}{C.~Di}, \bibinfo{author}{D.~Kolomenskiy},
  \bibinfo{author}{T.~Nakata}, \bibinfo{author}{H.~Liu},
\newblock \bibinfo{title}{Forewings match the formation of leading-edge
  vortices and dominate aerodynamic force production in revolving insect
  wings},
\newblock \bibinfo{journal}{Bioinspiration \& Biomimetics} \bibinfo{volume}{13}
  (\bibinfo{year}{2017}). \bibinfo{note}{Doi : \url{10.1088/1748-3190/aa94d7}}.
\bibitem[{van~de Meerendonk et~al.(2018)van~de Meerendonk, Percin, and van
  Oudheusden}]{FlexInfluenceMeerendonk}
\bibinfo{author}{R.~van~de Meerendonk}, \bibinfo{author}{M.~Percin},
  \bibinfo{author}{B.~W. van Oudheusden},
\newblock \bibinfo{title}{Three-dimensional flow and load characteristics of
  flexible revolving wings},
\newblock \bibinfo{journal}{Experiments in Fluids} \bibinfo{volume}{59}
  (\bibinfo{year}{2018}) \bibinfo{pages}{59:161}.
\bibitem[{Engels(2015)}]{ThomasThesis}
\bibinfo{author}{T.~Engels},
\newblock \bibinfo{title}{Numerical modeling of fluid-structure interaction in
  bio-inspired propulsion},
\newblock \bibinfo{journal}{PhD Thesis}  (\bibinfo{year}{2015}).
\bibitem[{Berger(1998)}]{BDFscheme}
\bibinfo{author}{J.~Berger},
\newblock \bibinfo{title}{A second order backward difference method with
  variable steps for a parabolic problem},
\newblock \bibinfo{journal}{BIT} \bibinfo{volume}{38} (\bibinfo{year}{1998})
  \bibinfo{pages}{644--662}.
\bibitem[{Combes and Daniel(2003)}]{CombesII}
\bibinfo{author}{S.~A. Combes}, \bibinfo{author}{T.~L. Daniel},
\newblock \bibinfo{title}{Flexural stiffness in insect wings {II}. {S}patial
  distribution and dynamic wing bending},
\newblock \bibinfo{journal}{The Journal of Experimental Biology}
  \bibinfo{volume}{206} (\bibinfo{year}{2003}) \bibinfo{pages}{2989--2997}.
\bibitem[{Walker et~al.(2009{\natexlab{a}})Walker, Thomas, and
  Taylor}]{WingDeformWalker1}
\bibinfo{author}{S.~M. Walker}, \bibinfo{author}{A.~L.~R. Thomas},
  \bibinfo{author}{G.~K. Taylor},
\newblock \bibinfo{title}{Deformable wing kinematics in the desert locust: how
  and why do camber, twist and topography vary through the stroke?},
\newblock \bibinfo{journal}{J. R. Soc. Interface} \bibinfo{volume}{6}
  (\bibinfo{year}{2009}{\natexlab{a}}) \bibinfo{pages}{735--747}.
\bibitem[{Walker et~al.(2009{\natexlab{b}})Walker, Thomas, and
  Taylor}]{WingDeformWalker2}
\bibinfo{author}{S.~M. Walker}, \bibinfo{author}{A.~L.~R. Thomas},
  \bibinfo{author}{G.~K. Taylor},
\newblock \bibinfo{title}{Photogrammetric reconstruction of high-resolution
  surface topographies and deformable wing kinematics of tethered locusts and
  free-flying hoverflies},
\newblock \bibinfo{journal}{J. R. Soc. Interface} \bibinfo{volume}{6}
  (\bibinfo{year}{2009}{\natexlab{b}}) \bibinfo{pages}{351–366}.
\bibitem[{Walker et~al.(2010)Walker, Thomas, and Taylor}]{WingDeformWalker3}
\bibinfo{author}{S.~M. Walker}, \bibinfo{author}{A.~L.~R. Thomas},
  \bibinfo{author}{G.~K. Taylor},
\newblock \bibinfo{title}{Deformable wing kinematics in free-flying
  hoverflies},
\newblock \bibinfo{journal}{J. R. Soc. Interface} \bibinfo{volume}{7}
  (\bibinfo{year}{2010}) \bibinfo{pages}{131–142}.
\bibitem[{Bomphrey et~al.(2005)Bomphrey, Lawson, Harding, Taylor, and
  R.}]{WingDeformBomphrey}
\bibinfo{author}{J.~R. Bomphrey}, \bibinfo{author}{N.~J. Lawson},
  \bibinfo{author}{N.~J. Harding}, \bibinfo{author}{G.~K. Taylor},
  \bibinfo{author}{T.~A.~L. R.},
\newblock \bibinfo{title}{The aerodynamics of manduca sexta: digital particle
  image velocimetry analysis of leading-edge vortex},
\newblock \bibinfo{journal}{J. Exp. Biol} \bibinfo{volume}{208}
  (\bibinfo{year}{2005}) \bibinfo{pages}{1079--1094}.
\bibitem[{Mountcastle and Daniel(2009)}]{WingDeformMountcastle}
\bibinfo{author}{A.~M. Mountcastle}, \bibinfo{author}{T.~L. Daniel},
\newblock \bibinfo{title}{Aerodynamic and functional consequences of wing
  compliance},
\newblock \bibinfo{journal}{Exp. Fluids} \bibinfo{volume}{46}
  (\bibinfo{year}{2009}) \bibinfo{pages}{873--882}.
\bibitem[{Fenner(1986)}]{NonplanarMembraneFenner}
\bibinfo{author}{R.~T. Fenner}, \bibinfo{title}{Engineering Elasticity:
  Application of Numerical and Analytical Techniques}, \bibinfo{publisher}{New
  York: John Wiley}, \bibinfo{year}{1986}.
\bibitem[{White(1985)}]{NonplanarMembraneWhite}
\bibinfo{author}{R.~E. White}, \bibinfo{title}{An Introduction to the Finite
  Element Method with Applications to Nonlinear Problems},
  \bibinfo{publisher}{New York: John Wiley}, \bibinfo{year}{1985}.
\bibitem[{Chen and Boyle(2014)}]{MembraneBoyle}
\bibinfo{author}{M.~Chen}, \bibinfo{author}{F.~Boyle},
\newblock \bibinfo{title}{Investigation of membrane mechanics using spring
  networks: application to red-blood-cell modelling},
\newblock \bibinfo{journal}{Materials Science and Engineering}
  \bibinfo{volume}{43} (\bibinfo{year}{2014}) \bibinfo{pages}{506--516}.
\bibitem[{Omori et~al.(2011)Omori, Ishikawa, Barth\`es-Biesel, Salsac, Walter,
  Imai, and Yamaguchi}]{MembraneOmori}
\bibinfo{author}{T.~Omori}, \bibinfo{author}{T.~Ishikawa},
  \bibinfo{author}{D.~Barth\`es-Biesel}, \bibinfo{author}{A.-V. Salsac},
  \bibinfo{author}{J.~Walter}, \bibinfo{author}{Y.~Imai},
  \bibinfo{author}{T.~Yamaguchi},
\newblock \bibinfo{title}{Comparison between spring network models and
  continuum constitutive laws: Application to the large deformation of a
  capsule in shear flow},
\newblock \bibinfo{journal}{Physical Review E} \bibinfo{volume}{83}
  (\bibinfo{year}{2011}) \bibinfo{pages}{041918}.
\bibitem[{Deussen et~al.(1995)Deussen, Kobbelt, and Tucke}]{MembraneDeussen}
\bibinfo{author}{O.~Deussen}, \bibinfo{author}{L.~Kobbelt},
  \bibinfo{author}{P.~Tucke},
\newblock \bibinfo{title}{Using simulated annealing to obtain good nodal
  approximations of deformable bodies},
\newblock \bibinfo{journal}{Springer-Verlag}  (\bibinfo{year}{1995})
  \bibinfo{pages}{30--43}.
\bibitem[{Lloyd et~al.(2007)Lloyd, Sz{\'e}kely, and Harders}]{MSMLloyd}
\bibinfo{author}{B.~A. Lloyd}, \bibinfo{author}{G.~Sz{\'e}kely},
  \bibinfo{author}{M.~Harders},
\newblock \bibinfo{title}{Identification of spring parameters for deformable
  object simulation},
\newblock \bibinfo{journal}{IEEE Transactions on Visualization and Computer
  Graphics} \bibinfo{volume}{13} (\bibinfo{year}{2007}).
\bibitem[{Louchet et~al.(1995)Louchet, Provot, and
  Crochemore}]{MembraneLouchet}
\bibinfo{author}{J.~Louchet}, \bibinfo{author}{X.~Provot},
  \bibinfo{author}{D.~Crochemore},
\newblock \bibinfo{title}{Evolutionary identification of cloth animation
  models},
\newblock \bibinfo{journal}{Springer-Verlag}  (\bibinfo{year}{1995})
  \bibinfo{pages}{44--54}.
\bibitem[{Bianchi et~al.(2003)Bianchi, Harders, and Sz\'ekely}]{MSMBianchiMesh}
\bibinfo{author}{G.~Bianchi}, \bibinfo{author}{M.~Harders},
  \bibinfo{author}{G.~Sz\'ekely},
\newblock \bibinfo{title}{Mesh topology identification for mass-spring models},
\newblock \bibinfo{journal}{Proc. Medical Image Computing and Computer-Assisted
  Intervention (MICCAI ’03)} \bibinfo{volume}{1} (\bibinfo{year}{2003}).
\bibitem[{Bianchi et~al.(2004)Bianchi, Solenthaler, Sz\'ekely, and
  Harders}]{MSMBianchiParam}
\bibinfo{author}{G.~Bianchi}, \bibinfo{author}{B.~Solenthaler},
  \bibinfo{author}{G.~Sz\'ekely}, \bibinfo{author}{M.~Harders},
\newblock \bibinfo{title}{Simultaneous topology and stiffness identification
  for mass-spring models based on fem reference deformations},
\newblock \bibinfo{journal}{Proc. Medical Image Computing and Computer-Assisted
  Intervention (MICCAI ’04)} \bibinfo{volume}{2} (\bibinfo{year}{2004}).
\bibitem[{Logan(2010)}]{FEMLogan}
\bibinfo{author}{D.~L. Logan}, \bibinfo{title}{A first course in the Finite
  Element Method, 5th Revised edition}, \bibinfo{publisher}{CL Engineering},
  \bibinfo{year}{2010}.
\bibitem[{Barten(1944)}]{BeamDeflection}
\bibinfo{author}{H.~J. Barten},
\newblock \bibinfo{title}{On the deflection of a cantilever beam},
\newblock \bibinfo{journal}{Quarterly of Applied Mathematics}
  \bibinfo{volume}{2} (\bibinfo{year}{1944}) \bibinfo{pages}{168--171}.
\bibitem[{Bisshopp and Drucker(1945)}]{BeamLargeDeflectionSolution}
\bibinfo{author}{K.~E. Bisshopp}, \bibinfo{author}{D.~C. Drucker},
\newblock \bibinfo{title}{Large deflections of cantilever beams},
\newblock \bibinfo{journal}{Quart. Appl. Math}  (\bibinfo{year}{1945})
  \bibinfo{pages}{272--275}.
\bibitem[{Wang et~al.(2006)Wang, Chen, and Liao}]{BeamLargeDeflection}
\bibinfo{author}{J.~Wang}, \bibinfo{author}{J.~K. Chen},
  \bibinfo{author}{S.~Liao},
\newblock \bibinfo{title}{An explicit solution of the large deformation of a
  cantilever beam under point load at the free tip},
\newblock \bibinfo{journal}{J. Comput. Appl. Math.} \bibinfo{volume}{212}
  (\bibinfo{year}{2006}).
\bibitem[{Engels et~al.(2017)Engels, Kolomenskiy, Schneider, and
  Sesterhenn}]{ThomasElasSwimmer}
\bibinfo{author}{T.~Engels}, \bibinfo{author}{D.~Kolomenskiy},
  \bibinfo{author}{K.~Schneider}, \bibinfo{author}{J.~Sesterhenn},
\newblock \bibinfo{title}{Numerical simulation of vortex-induced drag of
  elastic swimmer models},
\newblock \bibinfo{journal}{Theoretical and Applied Mechanics Letters}
  \bibinfo{volume}{27} (\bibinfo{year}{2017}) \bibinfo{pages}{280--285}.
\bibitem[{Kolomenskiy et~al.(2019)Kolomenskiy, Ravi, Xu, Ueyama, Jakobi,
  Engels, Nakata, Sesterhenn, Farge, Schneider, Onishi, and
  Liu}]{BumblebeeWingStructure}
\bibinfo{author}{D.~Kolomenskiy}, \bibinfo{author}{S.~Ravi},
  \bibinfo{author}{R.~Xu}, \bibinfo{author}{K.~Ueyama},
  \bibinfo{author}{T.~Jakobi}, \bibinfo{author}{T.~Engels},
  \bibinfo{author}{T.~Nakata}, \bibinfo{author}{J.~Sesterhenn},
  \bibinfo{author}{M.~Farge}, \bibinfo{author}{K.~Schneider},
  \bibinfo{author}{R.~Onishi}, \bibinfo{author}{H.~Liu},
\newblock \bibinfo{title}{The dynamics of passive feathering rotation in
  hovering flight of bumblebees.},
\newblock \bibinfo{journal}{J. Fluids. Struc.}  (\bibinfo{year}{2019}).
  \bibinfo{note}{Doi: \url{10.1016/j.jfluidstructs.2019.03.021}}.
\bibitem[{Vincent and Wegst(2004)}]{CuticleProperties}
\bibinfo{author}{J.~F.~V. Vincent}, \bibinfo{author}{U.~G.~K. Wegst},
\newblock \bibinfo{title}{Design and mechanical properties of insect cuticle},
\newblock \bibinfo{journal}{Arthropod Structure and Development}
  \bibinfo{volume}{33} (\bibinfo{year}{2004}) \bibinfo{pages}{187--199}.
\bibitem[{Angot et~al.(1999)Angot, Bruneau, and Fabrie}]{VolPenaAngot}
\bibinfo{author}{P.~Angot}, \bibinfo{author}{C.~Bruneau},
  \bibinfo{author}{P.~Fabrie},
\newblock \bibinfo{title}{A penalization method to take into account obstacles
  in incompressible viscous flows},
\newblock \bibinfo{journal}{Numer. Math.}  (\bibinfo{year}{1999})
  \bibinfo{pages}{81:497--520}.
\bibitem[{Kolomenskiy and Schneider(2009)}]{VolPenaDmitry}
\bibinfo{author}{D.~Kolomenskiy}, \bibinfo{author}{K.~Schneider},
\newblock \bibinfo{title}{A {F}ourier spectral method for the
  {N}avier--{S}tokes equations with volume penalization for moving solid
  obstacles},
\newblock \bibinfo{journal}{J. Comput. Phys.}  (\bibinfo{year}{2009})
  \bibinfo{pages}{228:5687--5709}.
\bibitem[{Eberly(1999)}]{TriDistance}
\bibinfo{author}{D.~Eberly},
\newblock \bibinfo{title}{Distance between point and triangle in 3d},
\newblock \bibinfo{journal}{Geometric Tools, LLC}  (\bibinfo{year}{1999}).
  \bibinfo{note}{{h}ttp://www.magic-software.com/Documentation/pt3tri3.pdf.}
\bibitem[{Yang et~al.(2009)Yang, Zhang, Li, and He}]{DeltaInterp}
\bibinfo{author}{X.~Yang}, \bibinfo{author}{X.~Zhang}, \bibinfo{author}{Z.~Li},
  \bibinfo{author}{G.-W. He},
\newblock \bibinfo{title}{A smoothing technique for discrete delta functions
  with application to immersed boundary method in moving boundary simulations},
\newblock \bibinfo{journal}{Journal of Computational Physics}
  \bibinfo{volume}{228} (\bibinfo{year}{2009}) \bibinfo{pages}{7821--7836}.
\bibitem[{Turek and Hron(2006)}]{Turek1}
\bibinfo{author}{S.~Turek}, \bibinfo{author}{J.~Hron},
\newblock \bibinfo{title}{Proposal for numerical benchmarking of
  fluid-structure interaction between an elastic object and laminar
  incompressible flow. {I}n: H.{J}. {B}ungartz and {M}. {S}ch\"afer (eds)
  {F}luid-{S}tructure {I}nteraction. {L}ecture {N}otes in {C}omputational
  {S}cience and {E}ngineering},
\newblock \bibinfo{journal}{Springer Berlin Heidelberg} \bibinfo{volume}{53}
  (\bibinfo{year}{2006}) \bibinfo{pages}{371--385}.
\bibitem[{Turek et~al.(2010)Turek, Hron, Razzaq, Wobker, and
  Sch\"afer}]{Turek2}
\bibinfo{author}{S.~Turek}, \bibinfo{author}{J.~Hron},
  \bibinfo{author}{M.~Razzaq}, \bibinfo{author}{H.~Wobker},
  \bibinfo{author}{M.~Sch\"afer},
\newblock \bibinfo{title}{Numerical benchmarking of fluid-structure
  interaction: A comparison of different discretization and solution
  approaches. {I}n: H.{J}. {B}ungartz, {M}. {M}ehl and {M}. {S}ch\"afer (eds)
  {F}luid-{S}tructure {I}nteraction {II}. {L}ecture {N}otes in {C}omputational
  {S}cience and {E}ngineering},
\newblock \bibinfo{journal}{Springer Berlin Heidelberg} \bibinfo{volume}{73}
  (\bibinfo{year}{2010}) \bibinfo{pages}{413--424}.
\bibitem[{Zhao et~al.(2009)Zhao, Huang, Deng, and Sane}]{FlexInfluenceZhao}
\bibinfo{author}{L.~Zhao}, \bibinfo{author}{Q.~Huang},
  \bibinfo{author}{X.~Deng}, \bibinfo{author}{S.~P. Sane},
\newblock \bibinfo{title}{Aerodynamic effects of flexibility in flapping
  wings},
\newblock \bibinfo{journal}{J R Soc Interface} \bibinfo{volume}{7}
  (\bibinfo{year}{2009}) \bibinfo{pages}{485--497}.

\end{thebibliography}


\end{document}